\documentclass[]{emulateapj}
\usepackage[]{natbib}
\usepackage{times}

\pdfoutput=0

\shorttitle{Accretion onto IMBHs Regulated by Radiative Feedback}
\shortauthors{Park and Ricotti}

\received{}
\accepted{}
\begin{document}

\title{Accretion onto Intermediate Mass Black Holes Regulated by
  Radiative Feedback.I\\ Parametric Study for Spherically Symmetric
  Accretion}

\author{KwangHo Park} \affil{Department of Astronomy, University of
Maryland, College Park, MD 20740} \email{kpark@astro.umd.edu}

\author{Massimo Ricotti} \affil{Department of Astronomy, University of
Maryland, College Park, MD 20740} \email{ricotti@astro.umd.edu}

\begin{abstract}
  We study the effect of radiative feedback on accretion onto
  intermediate mass black holes (IMBHs) using the hydrodynamical code
  ZEUS-MP with a radiative transfer algorithm. In this paper, the
  first of a series, we assume accretion from a uniformly dense gas
  with zero angular momentum and extremely low metallicity. Our 1D and
  2D simulations explore how X-ray and UV radiation emitted near the
  black hole regulates the gas supply from large scales.  Both 1D and
  2D simulations show similar accretion rate and period between peaks
  in accretion, meaning that the hydro-instabilities that develop in
  2D simulations do not affect the mean flow properties. We present a
  suite of simulations exploring accretion across a large parameter
  space, including different radiative efficiencies and radiation
  spectra, black hole masses, density and temperature, $T_\infty$, of
  the neighboring gas. In agreement with previous studies we find
  regular oscillatory behavior of the accretion rate, with duty cycle
  $\sim 6\%$, mean accretion rate 3\% $(T_{\infty}/10^4~{\rm
    K})^{2.5}$ of the Bondi rate and peak accretion $\sim 10$ times
    the mean for $T_{\infty}$ ranging between $3000~$K and $15000~$K. We
  derive parametric formulas for the period between bursts, the mean
  accretion rate and the peak luminosity of the bursts and thus
  provide a formulation of how feedback regulated accretion
  operates. The temperature profile of the hot ionized gas is crucial
  in determining the accretion rate, while the period of the bursts is
  proportional to the mean size of the Str\"{o}mgren sphere and we find
 qualitatively different modes of accretion in the high vs. low density
regimes. We also find that softer spectrum of radiation produces higher
mean accretion rate. However, it is still unclear what is the effect of
a significant time delay between the accretion rate at our inner boundary
and the output luminosity. Such a delay is expected in realistic cases with
  non-zero angular momentum and may affect the
  time-dependent phenomenology presented here. This study is a first
  step to model the growth of seed black holes in the early universe
  and to make a prediction of the number and the luminosity of
  ultra-luminous X-ray sources in galaxies produced by IMBHs accreting
  from the interstellar medium.
\end{abstract}



\section{INTRODUCTION}
The occurrence of gas accretion onto compact gravitating sources is
ubiquitous in the universe. The Bondi accretion formula
\citep{BondiH:44,Bondi:52}, despite the simplifying assumption of
spherical symmetry, provides a fundamental tool for understanding the
basic physics of the accretion process. Angular momentum of accreted
gas, in nearly all realistic cases, leads to the formation of an
accretion disk on scales comparable to or possibly much greater than
the gravitational radius of the black hole, $r_g \sim GM/c^2$, thus
breaking the assumption of spherical symmetry in the Bondi
solution. However, the fueling of the disk from scales larger than the
circularization radius $r_c \sim j^2/GM$, where $j$ is the gas
specific angular momentum, can be approximated by a quasi-radial
inflow. Thus, assuming that numerical simulations resolve the sonic
radius, $r_s$, the resolved gas flow is quasi-spherical if $r_c \ll
r_s$. The Bondi formula, which links the accretion rate to the
properties of the environment, such as the gas density and
temperature, or Eddington-limited rate are often used in cosmological
simulations to model the supply of gas to the accretion disk from
galactic scales \citep*{Volonteri:05,DiMatteo:08,
  Pelupessy:07,Greif:08,AlvarezWA:09}.

However, the Bondi formula is a crude estimation of the rate of gas
supply to the accretion disk because it does not take into account the
effect of accretion feedback loops on the surrounding environment.
Radiation emitted by black holes originates from gravitational
potential energy of inflowing gas \citep{Shapiro:73} and a substantial
amount of work has been performed to understand the simplest case of
spherical accretion onto compact X-ray sources or quasars. Several
authors have used hydrodynamical simulations to explore how feedback
loops operates and whether they produce time-dependent or a steady
accretion flows. A variety of feedback processes have been considered:
X-ray preheating, gas cooling, photo-heating and radiation pressure
\citep*{OstrikerWYM:76,CowieOS:78,BB:80,KrolikL:83,Vitello:84,WandelYM:84,
  MiloBCO:09,NovakOC:10, OstrikerCCNP:10}. Typically, the dominance of
one process over the others depends on the black hole mass and the
properties of the gas accreted by the black hole.  The qualitative
description of the problem is simple: gravitational potential energy
is converted into other forms of energy such as UV and X-ray photons
or jets, which act to reduce and reverse the gas inflow, either by
heating the gas or by producing momentum driven outflows
\citep*{CiottiO:07, CiottiOP:09,Proga:07,ProgaOK:08}.  In general
these feedback processes reduce the accretion rate and thus the
luminosity of the accreting black hole
\citep*{OstrikerWYM:76,Begelman:85,Ricotti:08}.  Consequently, the
time averaged accretion rate differs from Bondi's solution.  There
have been works on self-regulated accretion of supermassive black
holes (SMBHs) at the center of elliptical galaxies
\citep{Sazonov:05,CiottiO:07, CiottiOP:09} and radiation-driven
axisymmetric outflow in active galactic nuclei
\citep*{Proga:07,ProgaOK:08,KurosawaPN:09,KurosawaP:09a,KurosawaP:09b}.
However, far less has been done in order to quantify the simplest case
of spherical accretion onto IMBHs as a function of the properties of
the environment.  Recent theoretical \citep*[][hereafter
MBCO09]{MiloBCO:09} and numerical \citep*[][hereafter
MCB09]{MiloCB:09} works explore accretion of protogalactic gas onto
IMBHs in the first galaxies. MCB09 describes the accretion onto a
100~M$_\odot$ black hole from protogalactic gas of density
$n_{H,\infty}=10^7$~cm$^{-3}$ and temperature $T_\infty =10^4$~K. Our
study, which complements this recent numerical work, is a broader
investigation of accretion onto IMBHs for a set of several simulations
with a wide range of radiative efficiencies, black hole masses,
densities and sound speeds of the ambient gas. Our aim is to use
simulations to provide a physically motivated description of how
radiation modifies the Bondi solution and provide an analytical
formulation of the problem (see MBCO09).

The results of the present study will help to better understand the
accretion luminosities of IMBHs at high z and in the present-day
universe \citep*{Ricotti:09}. Applications of this work include
studies on the origin of ultra-luminous X-ray sources (ULXs)
\citep*{Krolik:81,Krolik:84,Ricotti:07, MackOR:07, Ricotti:08}, the
build up of an early X-ray background \citep*{VenkatesanGS:01,
RicottiO:04, MadauV:04, RicottiOG:05} and growth of SMBHs
\citep*{VolonteriHM:03,Volonteri:05,JohnsonB:07,Pelupessy:07,AlvarezWA:09}. For
example, different scenarios have been proposed for the formation of
quasars at $z \sim 6$ \citep{Fan:03}: growth by mergers, accretion
onto IMBHs, or direct formation of larger seed black holes from
collapse of quasi-stars \citep*{Carr:84,HaehneltNR:98,Fryer:01,
BegelmanVR:06,VolonteriLN:08,OmukaiSH:08,ReganH:09,MayerKEC:10} that
may form from metal free gas at the center of rare dark matter halos
\citep*{OhH:02}. Understanding the properties which determine the
efficiency of self-regulated accretion onto IMBHs is important to
estimate whether primordial black holes produced by Pop~III stars can
accrete fast enough to become SMBHs by redshift $z \sim 6$
\citep*{MadauR:01,VolonteriHM:03,YooM:04,Volonteri:05,JohnsonB:07,
Pelupessy:07,AlvarezWA:09}.


In this paper we focus on simulating accretion onto IMBH regulated by
photo-heating feedback in 1D and 2D hydrodynamic simulations, assuming
spherically symmetric initial conditions. We provide fitting formulas
for the mean and peak accretion rates, and the period between
accretion rate bursts as a function of the parameters we explore,
including radiative efficiency, black hole mass, gas density,
temperature and spectrum of radiation. In \S~2 we introduce basic
concepts and definitions in the problem. Numerical procedures and
physical processes included in the simulations are discussed in \S~3.
Our simulation results and the parameter study are shown in \S~4. In
\S~5 we lay out a physically motivated model that describes the
results of the simulations. Finally, a summary and discussion are
given in \S~6.

\section{BASIC DEFINITIONS}
\subsection{Bondi Accretion and Eddington luminosity}

The assumption of spherical symmetry allows to treat accretion
problems analytically. The solution \citep{Bondi:52} provides the
typical length scale at which gravity affects gas dynamics and the
typical accretion rate as a function of the black hole mass $M_{bh}$,
ambient gas density $\rho_\infty$ and sound speed $c_{s,\infty}$. The
Bondi accretion rate for a black hole at rest is
\begin{eqnarray}
\dot{M}_B &=& 4\pi\lambda_{B} r_b^2 \rho_{\infty} c_{s,\infty} \nonumber\\
   & = & 4 \pi \lambda_{B} \rho_\infty \frac{G^2 M_{bh}^2 }{c_{s,\infty}^3}, 
\end{eqnarray}
where $r_b = GM_{bh}c_{s,\infty}^{-2}$ is the Bondi radius, and
$\lambda_B$ is the dimensionless mass accretion rate, which depends on
the polytropic index, $\gamma$, of the gas equation of state
$P=K\rho^\gamma$:
\begin{eqnarray}
\lambda_{B} &=& \frac{1}{4} \left[ \frac{2}{5-3\gamma} \right]^{\frac{5-3\gamma}{2(\gamma -1)}}.
\end{eqnarray}
The value of $\lambda_{B}$ ranges from $e^{3/2}/4 \simeq 1.12$ for an
isothermal gas ($\gamma = 1$ ) to $1/4$ for an adiabatic gas
($\gamma=5/3$).

However, a fraction of the gravitational potential energy of the
inflowing gas is necessarily converted into radiation or mechanical
energy when it approaches the black hole, significantly affecting the
accretion process. Photons emitted near the black hole heat and ionize
nearby gas, creating a hot bubble which exerts pressure on the
inflowing gas. Radiation pressure may also be important in reducing
the rate of gas inflow (see MBCO09). These processes may act as
self-regulating mechanisms limiting gas supply to the disk from larger
scales and, thus, controlling the luminosity of the black hole. We
quantify the reduction of the accretion rate with respect to the case
without radiative feedback by defining the dimensionless accretion
rate
\begin{eqnarray}
\lambda_{rad} \equiv \frac{\dot{M}}{\dot{M}_B}, 
\end{eqnarray}
where $\dot{M}_B$ is the Bondi accretion rate for isothermal gas (
$\dot{M}_B = e^{3/2} \pi G^2 M_{bh}^2 \rho_{\infty}
c_{s,\infty}^{-3}$). This definition of $\lambda_{rad}$ is consistent
with the one adopted by MBCO09.

\subsection{Luminosity and Radiative efficiency}

The Eddington luminosity sets an upper limit on the luminosity of a
black hole. In this limit the inward gravitational force on the gas
equals the radiation pressure from photons interacting with electrons
via Compton scattering. Although this limit can be evaded in some
special cases, observations suggest that black hole and SMBH
luminosity is sub-Eddington. The Eddington luminosity is thus,
\begin{eqnarray}
L_{Edd} &=& \frac{4\pi GM_{bh}m_{p}c}{\sigma_{T}} \simeq 3.3\times
10^6 L_{\sun} \left( \frac{M_{bh}}{100~M_{\sun}} \right).
\end{eqnarray}
The luminosity of an accreting black hole is related to the accretion
rate via the radiative efficiency $\eta$: $L =\eta \dot{M}c^2$.
From the Eddington luminosity, we define the Eddington gas accretion
rate $\dot{M}_{Edd} \equiv L_{Edd}c^{-2} $, and 
the dimensionless accretion rate and luminosity as
\begin{eqnarray}
\dot{m} \equiv \frac{\dot{M}}{\dot{M}_{Edd}}~~~ {\rm and}~~~
l \equiv \frac{L}{L_{Edd}}.
\end{eqnarray} 
Hence, in dimensionless units, the bolometric luminosity of the black
hole is $l=\eta \dot{m}$, where $\dot{m}$ is the accretion rate onto
the black hole. Note, that our definition of $\dot{M}_{Edd}$ is
independent of the radiative efficiency $\eta$. Therefore, if we
impose sub-Eddington luminosity of the black hole, the dimensionless
accretion rate ranges between $0 < \dot{m} \leq \frac{1}{\eta} $. The
radiative efficiency, $\eta$, depends on the geometry of the accretion
disk and on $\dot{m}$. For a thin disk, $\eta \simeq 0.1$, whereas
$\eta \propto \dot{m}$ for an advection dominated thick disk or for
spherical accretion \citep{Shapiro:73,Park:01}. In this study we
consider two idealized cases for the radiative efficiency. The case of
constant radiative efficiency $\eta=const$; and the case in which the
radiative efficiency has a dependence on the dimensionless accretion
rate and luminosity: $\eta=const$ for $l \ge 0.1$ and $\eta \propto
\dot{m}$ for $l < 0.1$. The second case we explored accounts for the lower
radiative efficiency expected when the accreted gas does not settle
into a thin disk. In both formulations the the radiative efficiency is
one of the free parameter we allow to vary and we do not find
important differences between the two cases. Observations of Sgr~A*,
the best studied case of low accretion rate onto a SMBH, suggest that
the radiative efficiency is indeed low but not as low as implied by
the scaling $\eta \propto \dot{m}$. Recent theoretical work by
\citep{Sharma:07} demonstrates that there is indeed a floor on the
radiative efficiency.


Because the Bondi rate, $\dot{M}_B$ does not include radiation
feedback effect, it provides an upper limit on the accretion rate from
large scales to radii near the black hole. The Eddington rate provides
the maximum accretion rate onto the black hole, limited by radiation
feedback at small radii. Thus, numerical simulations are necessary to
obtain realistic estimates of the accretion rates. If the accretion
rate onto the black hole is lower than the gas accretion from large
scales, the accreted material accumulates near the black hole,
creating a disk whose mass grows with time. We cannot simulate such a
scenario because it is too computationally challenging to resolve a
range of scales from the Bondi radius to the accretion disk in the
same simulations. Here we assume that accretion onto the black hole is
not limited by physical processes taking place on radial distances
much smaller than the sonic radius. For instance, even if angular
momentum of accreted gas is small and the circularization radius
$r_c\ll r_s$, further inflow will be slowed down with respect to the
free-fall rate. The rate of inflow will be controlled by angular
momentum loss (e.g.\, torques due to MHD turbulence) and there will be a
delay between the accretion rate at the inner boundary of our
simulation ($r_{min}$) and the accretion luminosity associated with
it. The effect of the aforementioned time delay on the feedback loop
is not considered in this paper but will be considered in future
works. We also assume that the effect of self-gravity is negligible in
our simulations since we have estimate that the mass within the HII
region around the black hole is smaller than the black hole mass for
$M_{bh}< 1000$~M$_\odot$.

If the rate of gas supply to the disk is given by the Bondi rate,
accretion onto the black hole is sub-Eddington for black hole masses
\begin{equation}
M_{bh} < {c_{s,\infty}^3 \over G n_{H,\infty} \sigma_T c \eta} \sim ~
40M_\odot~ T_{\infty,4}^{1.5} n_{H,5}^{-1} \eta_{-1}^{-1},
\label{eq:edd}
\end{equation}
where we use the notations of $T_{\infty,4} \equiv T_{\infty}/(10^4~{\rm K})$,
$n_{H,5}\equiv n_{H,\infty}/(10^5$~cm$^{-3})$ and
$\eta_{-1}\equiv\eta/10^{-1}$. Thus, in this regime we may assume that the
accretion is quasi-steady in the sense that the mean accretion rate
onto the black hole equals the gas supply from large scales when the
accretion rate is averaged over a sufficiently long time scale.



\section{NUMERICAL SIMULATIONS}
\subsection{ZEUS-MP and Radiative Transfer Module}

We perform a set of hydrodynamic simulations to understand accretion
onto IMBHs regulated by radiation feedback. Numerical simulations of
radiative feedback by black holes are challenging because they involve
resolving a large dynamical range in length scales. In this study we
use ZEUS-MP \citep{Hayes:06}, a modified parallel version of the
non-relativistic hydrodynamics code ZEUS \citep*{StoneN:92}. For the
present work we add a radiative transfer module \citep*{RicottiGS:01}
to ZEUS-MP to simulate radiative transfer of UV and X-ray ionizing
photons emitted near the black hole. A detailed description of the
numerical methods used to solve radiative transfer and tests of the
code are presented in the Appendix.
 
As X-ray and UV photons ionize the surrounding medium, different
reactions take place depending on the density and composition of the
gas. Photo-ionization changes the ionization fraction of H and He. The
detailed evolution of the Str\"{o}mgren sphere depends on the cooling
function $\Lambda(T,Z)$ of the gas and thus on the metallicity, $Z$,
and the fraction of gas in the molecular phase.  For a gas of
primordial composition, the cooling rate depends on the formation
rates of $H^{-}$ and $H_2$, which depend on both the redshift and the
intensity of the local dissociating background in the $H_2$
Lyman-Werner bands
\citep*[e.g.][]{ShapiroK:87,AbelAnninos:98,RicottiGS:02a,RicottiGS:02b}. In
addition, the cooling function may depend on redshift due to Compton
cooling of the electrons by CMB photons. We adopt atomic hydrogen
cooling for temperatures $T>10^4$~K, and use a simple parametric
function to model complicated cooling physics of gas at $T <
10^4~$K. Thus, the temperature structure inside the ionized bubble is
appropriate only for a low metallicity gas. For a subset of
simulations we also include the effect of helium photo-heating and
cooling. We assume that gas cooling at temperatures below $T_{\infty}$
is negligible in order to achieve thermal equilibrium in the initial
conditions far from the black hole. For the parameter space in which
we can neglect the effect of radiation pressure we find (see \S~5.1)
that the accretion rate is a function of the temperature both outside
and inside the HII region. The temperature outside the HII region
depends on the cooling function of gas at $T < 10^4~$K and on the
heating sources. The temperature inside the HII region depends on the
spectrum of radiation and cooling mechanism of gas at $T >
10^4$~K. Thus, it depends on the gas metallicity and the redshift at
which Compton cooling might become important. However, for the
parameter space we have explored we find that Compton cooling has a
minor effect on the temperature inside and outside the Str\"{o}mgren
sphere.

The gas heating rate depends on flux and spectral energy distribution
(SED) of the radiation emitted near the black hole. We assume a
luminosity of the black hole $l=\eta \dot{m}$ (see \S~2.2), where
$\dot{m}$ is calculated at the inner boundary in our simulation
(typically $r_{min}\sim 3\times10^{-5}$~pc). We adopt a single power law
$\nu^{-\alpha}$ for the SED, where the spectral index $\alpha$ is one
of the parameters we vary in our set of simulations.

We use an operator-split method to calculate the hydrodynamic step and
the radiative transfer and chemistry steps. The hydrodynamic
calculation is done using ZEUS-MP, then for the radiative transfer
calculation we use a ray tracing module \citep{RicottiGS:01}. The
radiative transfer module calculates chemistry, cooling and heating
processes along rays outgoing from the central black hole, and thus is
easily parallelized in the polar angle direction.

We perform 1D and 2D simulations in spherical coordinates. In both
cases we use a logarithmically spaced grid in the radial direction
typically with 256 to 512 cells to achieve high resolution near the
black hole. The size ratio between consecutive grids is chosen
according to the free parameters of the simulation to resolve the
ionization front and resolve the region where the gas is in free
fall. In the 2D simulations we use evenly spaced grids in the polar
angle direction and compute radiative transfer solutions in each
direction. Flow-out inner boundary conditions and flow-in outer
boundary conditions are used in the radial direction ($r$), whereas in
polar angle directions ($\theta$), reflective boundary conditions are
used.

To determine the optimal box size of the simulations we make sure that
we resolve important length scales in the problem: the inner Bondi
radius, $r_{b,in}$, the outer Bondi radius, $r_{b,\infty}$, the sonic
radius, $r_s$ and the ionization front, $R_s$. We select the value of
the inner boundary (typically ${\sim} 3\times 10^{-5}$~pc for
$M_{bh}=100~$M$_{\sun}$) to be smaller than the sonic point or the
inner Bondi radius (both still far larger than the Schwarzschild
radius of the black holes).  We find that once the sonic radius is
resolved, reducing the inner boundary box size does not create
significant differences in the results. In most cases the ionization
front is located outside of the outer Bondi radius and the box size is
selected to be large enough to cover both length scales. We select a
box size that achieves the highest possible resolution with a given
number of grids, making sure that the physical quantities around
boundaries remain constant during the simulations. The box is
sufficiently large to minimize the effect of spurious wave reflections
at the outer boundary.

In this paper, the first of a series, we adopt idealized initial
conditions of uniform density and temperature, zero velocity and zero
angular momentum of the gas relative to the black hole. In future work
we will relax some of these assumptions by adding turbulence in the
initial condition and considering the effect of black hole motion with
respect to the ambient medium and considering the effect of a
time-delay between the accretion rate at the inner boundary of our
simulations and the accretion luminosity. We assume monatomic,
non-relativistic ideal gas with $\gamma = 5/3$ which is initially
neutral (electron fraction $x_e \sim 10^{-5}$). In this paper we also
neglect the effect of radiation pressure. Our goal is to add to the
simulations one physical process at a time to understand which
feedback loop is dominant in a given subset of the parameter space. We
take this approach to attempt an interpretation of the simulation
results in the context of a physically motivated analytical
description of the accretion cycle. We will explore the effect of
radiation pressure due to HI ionization and Lyman-alpha scattering in
future works. However, a simple inspection of the relevant equations
suggests that radiation pressure is increasingly important for large
values of the ambient gas density ($n_{H,\infty} \sim 10^7$~cm$^{-3}$,
see MBCO09 and \S~6) since accretion rate approaches Eddington limit.

\begin{figure*}[thb]
\epsscale{0.9} \plotone{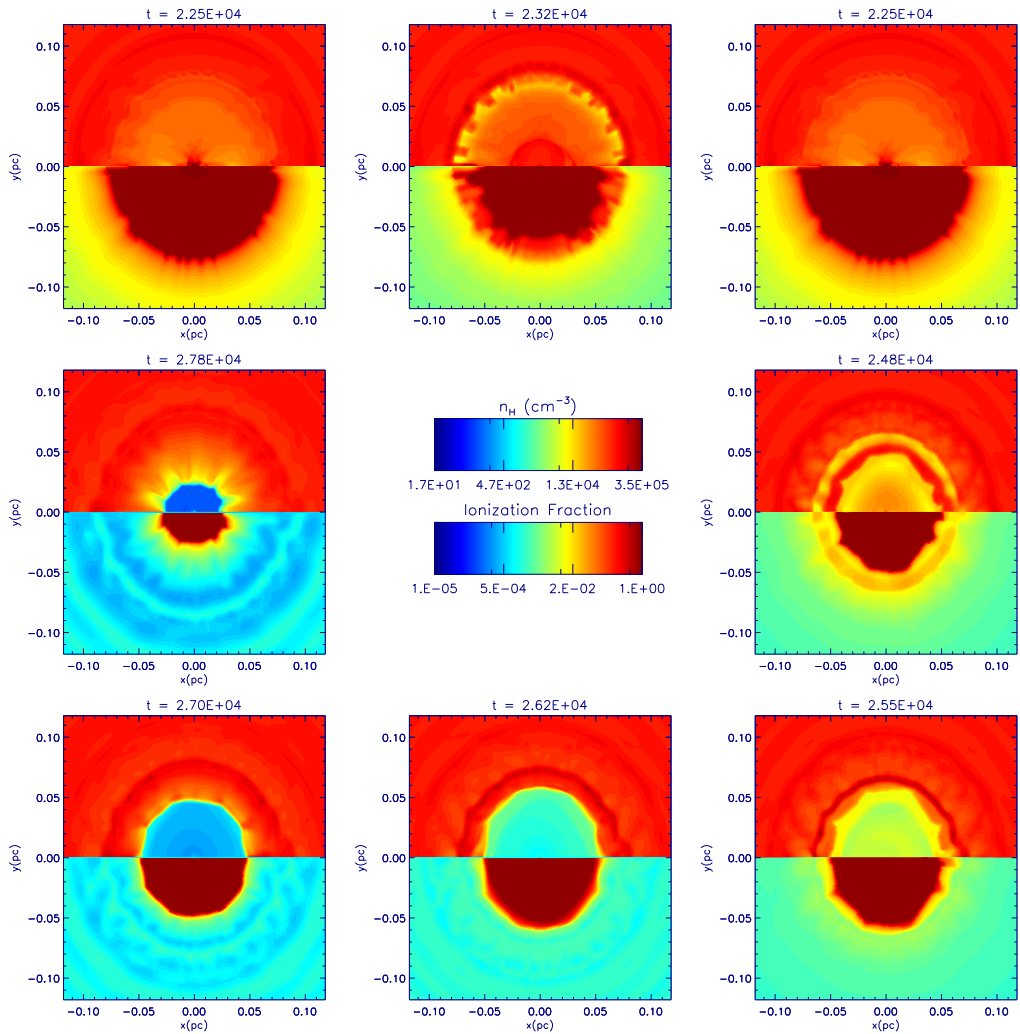}
\caption{Evolution of the gas density and ionization fraction in a
  simulation of an accreting black hole of mass
  $M_{bh}=100$~M$_{\odot}$, gas density $n_{H,\infty} = 10^5$
  cm$^{-3}$, and temperature $T_{\infty}=10^4$K. In each panel the top
  halves show the density (number of hydrogen atoms per cm$^3$) and
  the bottom halves show the ionization fraction, $x_e=n_e/n_H$, of
  the gas. The evolutionary sequences are shown in a clockwise
  direction. {\it Top panels from left to right }: A Str\"{o}mgren
  sphere forms fueled by ionizing photons as the black hole accretes
  gas. The higher pressure inside the Str\"{o}mgren sphere stops the
  gas inflow while the black hole at the center consumes the hot gas
  inside the ionization front. Inflowing gas accumulates in a dense
  shell outside the hot bubble while exponential decay of the
  accretion rate occurs due to decreasing density inside the hot
  bubble as gas depletion continues. Although the number of emitted
  ionizing photons decreases, the ionized sphere maintains its size
  because of the decrease in density inside the hot bubble. {\it
  Bottom panels from right to left}: The density of hot gas inside the
  Str\"{o}mgren sphere keeps decreasing until pressure equilibrium
  across the front can no longer be maintained. {\it Middle left}: The
  dense shell in front of the Str\"{o}mgren sphere collapses onto the
  black hole and this leads to a burst of accretion luminosity. {\it
  Top left}: The Str\"{o}mgren sphere reaches its maximum size and the
  simulation cycle repeats.}
\label{evolution}
\end{figure*}

\section{RESULTS}
\subsection{Qualitative description of accretion regulated by radiative feedback}

Our simulations show that UV and X-ray photons modify the thermal and
dynamical structure of the gas in the vicinity of the Bondi radius. A
hot bubble of gas is formed due to photo-heating by high energy
photons and sharp changes of physical properties such as density,
temperature, and ionization fraction occur at the ionization front.
Figure~\ref{evolution} shows 8 snapshots from one of our 2D
simulations. Top half of each snapshot shows the gas density and the
bottom half shows the hydrogen ionization fraction. We show the
periodic oscillation of the density and the ionization fraction from a
2D simulation in Figure~\ref{evolution}. The time evolution of the
density, temperature and ionization fraction profiles for the 1D
simulation are shown in Figure~\ref{profile}. We can identify 3
evolutionary phases that repeat cyclically:

\begin{figure}[thb]
\epsscale{0.5}
\plotone{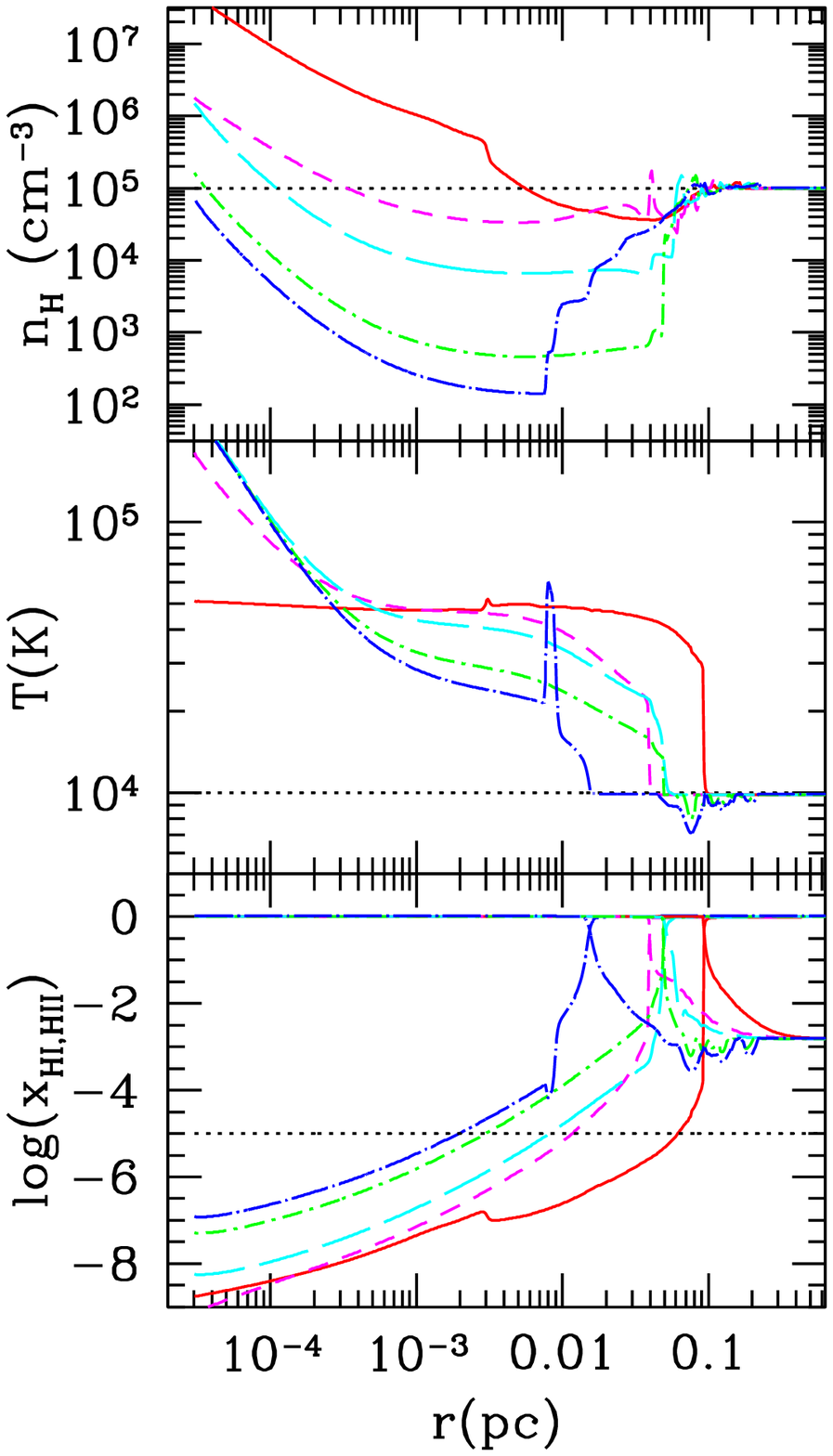}
\caption{{\it Top to bottom} : Radial profiles of density, temperature
  and neutral/ionization fractions in 1D simulation for $\eta$=0.1,
  $M_{bh}$=100~M$_{\sun}$, $n_{H,\infty}$=$10^5~$cm$^{-3}$ and
  $T_{\infty}$=$10^4~$K. Different lines indicates profiles at
  different times: t=0.0 (dotted), t=1.13$\times$$10^4$ (solid),
  t=1.28$\times$$10^4$ (short dashed), t=1.43$\times$$10^4$ (long
  dashed), t=1.58$\times$$10^4$ (dot-short dashed),
  t=1.71$\times$$10^4$~yr (dot-long dashed). {\it Solid lines} : at the
  maximum expansion of the Str\"{o}mgren sphere. {\it Dot-long dashed
    lines} : at the collapsing phase of dense shell. Physical
  properties inside the Str\"{o}mgren sphere change as a function of
  time. The number density and temperature of hydrogen decrease with
  time after the burst. The neutral fraction increases as a function
  of time from the burst.}
\label{profile}
\end{figure}

\begin{figure*}[thb]
\epsscale{1.0}
\plottwo{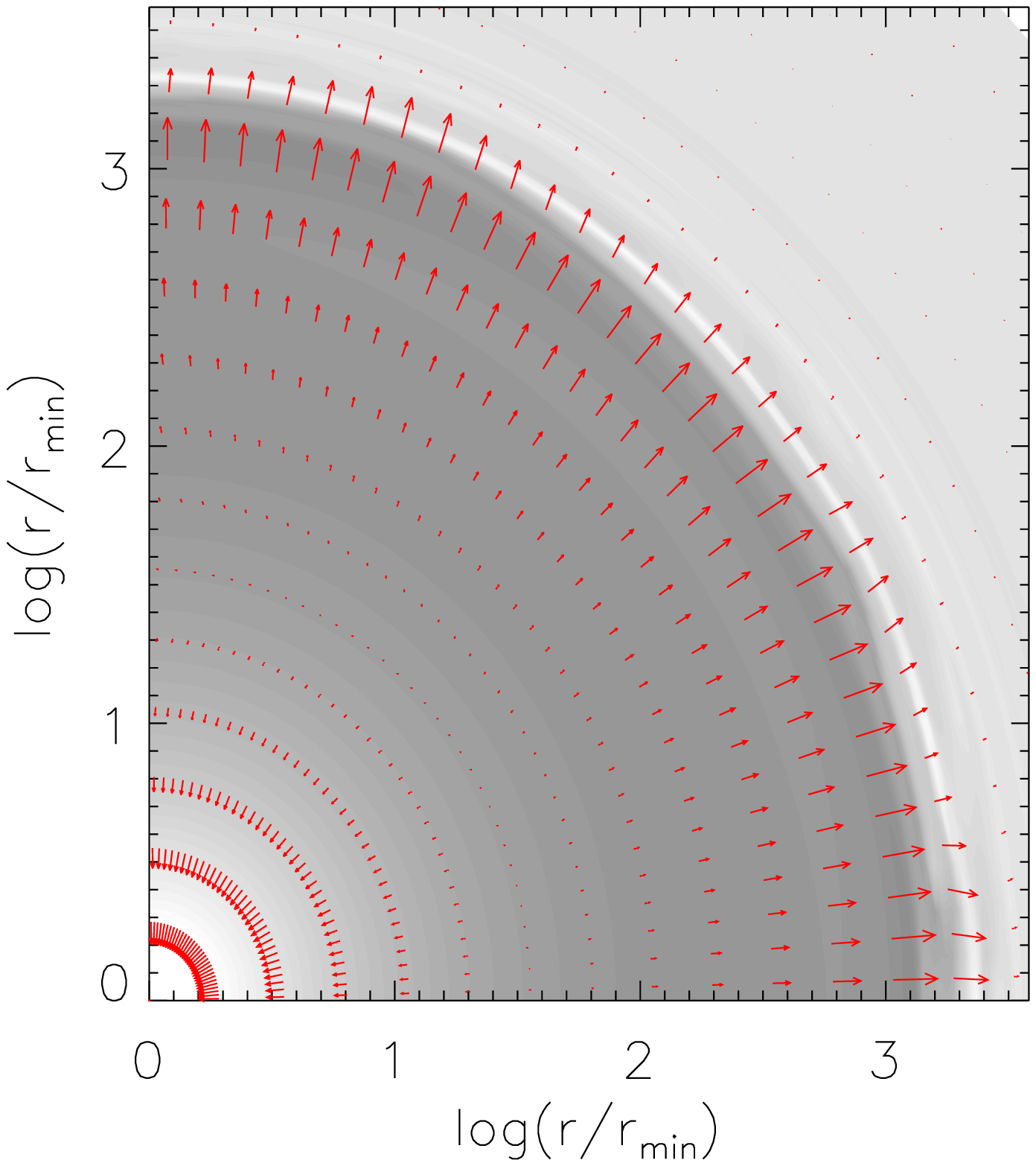}{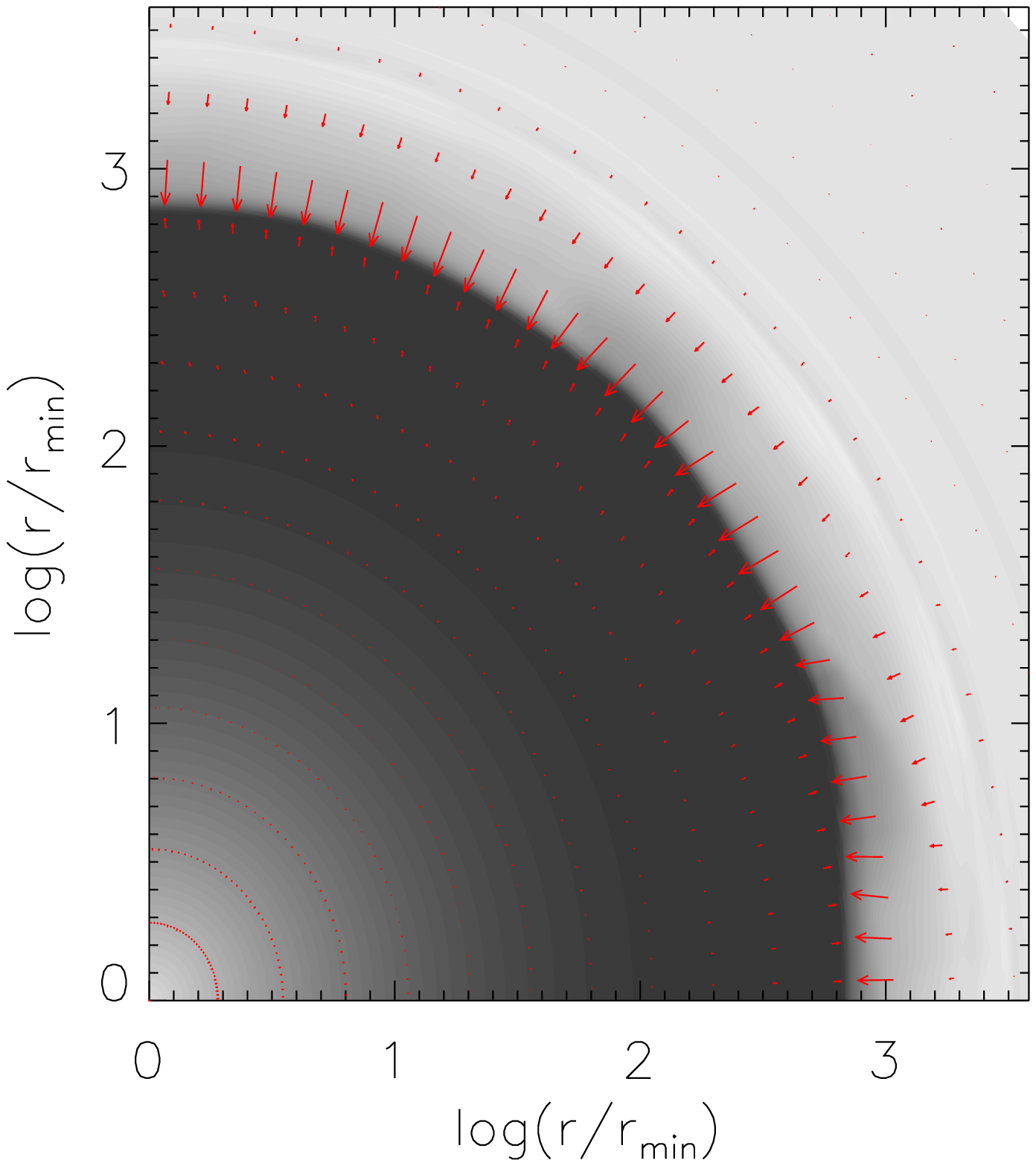}
\caption{Gas density and velocity field for the simulation with
  $\eta=0.1$, $M_{bh}=100$~M$_{\odot}$, $n_{H,\infty} = 10^5$
  cm$^{-3}$, and $T_{\infty}=10^4$K. {\it Left}: When a Str\"{o}mgren
  sphere is formed, gas inside the hot bubble is depleted by accretion
  onto the black hole and the outflow toward the dense shell due to
  pressure gradient. {\it Right}: Gas depletion inside the
  Str\"{o}mgren sphere leads to the collapse of the dense shell,
  creating a burst of accretion.}
\label{vel}
\end{figure*}

1) Once the Str\"{o}mgren sphere is formed, it expands and the gas
density inside that hot bubble decreases maintaining roughly pressure
equilibrium across the ionization front. At the front, gas inflow is
stopped by the hot gas and the average gas density inside the bubble
decreases due to the following two physical processes. First, the
black hole continues accreting hot gas within an accretion radius,
$r_{acc}$, defined as the radius where the gravitational force of the
black hole dominates the thermal energy of the hot gas. The accretion
radius is similar to the Bondi radius defined by the temperature
inside Str\"{o}mgren sphere, but there exists a difference between
them since the kinematic and thermal structure of gas is modified
significantly by the photo-heating and cooling. Second, the gas
between $r_{acc}$ and the ionization front moves towards the
ionization front due to pressure gradients. The left panel of Figure
\ref{vel} shows inflowing gas within $r_{acc}$ and outflowing gas
outside $r_{acc}$. A dense shell forms just outside the ionization
front. Thus, the mass of the shell grows because gravity pulls distant
gas into the system at the same time that gas within the the hot
bubble is pushed outwards.

\begin{figure}[thb]
\epsscale{1.0}
\plotone{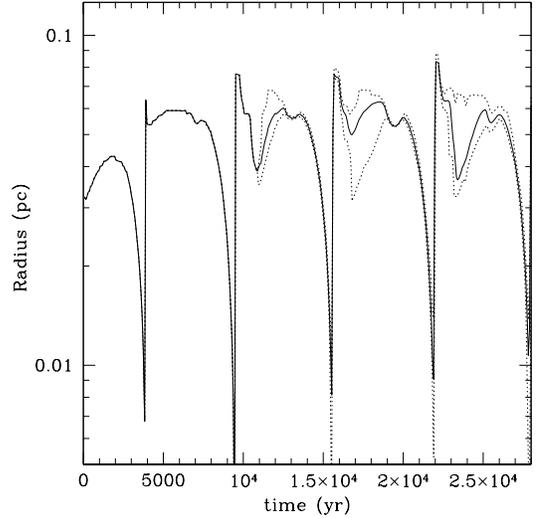}
\caption{ Evolution of Str\"{o}mgren radius with time for 2D
  simulation with $\eta=0.1$, $M_{bh} =100$~M$_\sun$,
  $n_{H,\infty}=10^5$~cm$^{-3}$ and $T_{\infty}= 10^4$~K. The solid
  line shows the mean size of the Str\"{o}mgren radius and dotted
  lines show the minimum and maximum Str\"{o}mgren radii. It shows the
  same period of oscillation seen in accretion rate as a function of
  time. In general, the Str\"{o}mgren radius is proportional to the
  accretion rate which determines the number of ionizing photons. When
  the accretion rate is maximum, the size of the Str\"{o}mgren sphere
  also has its maximum size.}

\label{ifront}
\end{figure}

2) As the average density inside the hot bubble decreases, the
accretion rate diminishes. During this process the radius of the
Str\"{o}mgren sphere remains approximately constant since the reduced
number of ionizing UV and X-ray photons is still sufficient to ionize
the rarefied hot bubble. Figure~\ref{ifront} illustrates this. Thus,
the average gas temperature, ionization fraction and the size of the
HII region remain constant. As the accretion rate increases during
the burst, it produces a rapid expansion of the Str\"{o}mgren sphere
radius. During one cycle of oscillation, there are small peaks in the
Str\"{o}mgren sphere radius which are associated with minor increases
in the accretion rate. Rayleigh-Taylor (RT) instabilities develop
quickly when the accretion rate increases. In these phases, the
acceleration of the dense shell is directed toward the black hole, so
the dense shell, supported by more rarefied gas, becomes RT unstable.

3) As gas depletion continues, the pressure inside the hot bubble
decreases to the point where equilibrium at the ionization front
breaks down. The outward pressure exerted by the hot bubble becomes
too weak to support the gravitational force exerted on the dense
shell. The dense shell of gas collapses toward the black hole,
increasing dramatically the accretion rate and creating a burst of
ionizing photons. The ionization front propagates outward in a
spherically symmetric manner, creating a large Str\"{o}mgren sphere
and returning to the state where the high pressure inside the
Str\"{o}mgren sphere suppresses gas inflow from outside.

\subsection{Comparison of 1D and 2D simulations}
In agreement with previous studies, our simulations show that
radiation feedback induces regular oscillations of the accretion rate
onto IMBH. This result is in good agreement with numerical work by
MCB09 for accretion onto a $100~$M$_\sun$ black hole from a high density
($n_{H,\infty}=10^7$~cm$^{-3}$) and high temperature
($T_{\infty}=10^4$~K) gas. Periodic oscillatory behavior is found in
all our simulations for different combinations of parameters, when
assuming spherically symmetric initial conditions and a stationary
black hole. This oscillation pattern is quite regular and no sign of
damping is observed for at least $\sim 10$ cycles.

\begin{figure}[thb]
\epsscale{1.0}
\plotone{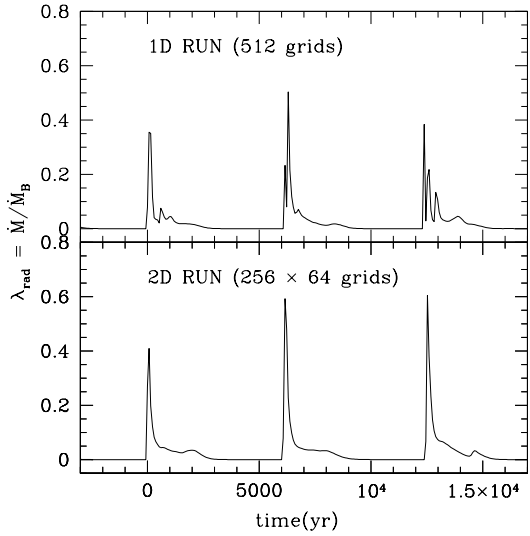}
\caption{Accretion rates as a function of time in 1D and 2D
  simulations with $\eta=0.1$, $M_{bh} =100$~M$_\sun$,
  $n_{H,\infty}=10^5$~cm$^{-3}$ and $T_{\infty}= 10^4$~K. Both results
  show similar oscillation patterns with the same period and average
  accretion rate.}
\label{1d2d}
\end{figure}

For the same parameters, our 1D and 2D simulations are nearly
identical in terms of oscillatory behavior in accretion rate and
Str\"{o}mgren sphere size. Figure~\ref{1d2d} shows accretion rate in
1D and 2D simulations for $M_{bh}$=100 M$_\sun$, $T_{\infty}=10^4$~K and
$n_{H,\infty}=10^5$~cm$^{-3}$. Note the similar pattern in accretion rate and
period between bursts. This indicates that the 1D result adequately
represents 2D cases when the accretion flow does not have significant
angular momentum.


Moreover, this result demonstrates that RT instabilities
which we observe in the 2D simulations do not affect the mean accretion rate
or the period of oscillations. The RT instability develops during the
phase when the dense shell in front of the ionization front is
supported against gravitational accretion by the low density medium
inside the hot bubble \citep{Whalen:08a,Whalen:08b}. The top
panels in Figure~\ref{evolution} show small instabilities when
ionization fronts move outward, which largely decay over
time. The pressure gradient inside the Str\"{o}mgren sphere creates an
outward force which helps suppress the development of the instability.

In summary, we believe that 1D simulations can be used in place of
higher dimension simulations to determine the cycle and magnitude of
the periodic burst of gas accretion onto IMBH. This allows us to
reduce the computational time required to explore a large range of
parameter space.

\subsection{ Parameter space exploration}

In this section we present the results of a set of 1D simulations
aimed at exploring the dependence of the accretion rate and the period
of oscillations of the black hole luminosity as a function of the
black hole mass, $M_{bh}$, the ambient gas density, $n_{H,\infty}$,
temperature, $T_\infty$, and the radiative efficiency $\eta$. In
$\S~$5 we present results in which we allow the spectrum of ionizing
radiation to vary as well. The accretion can be described by three
main parameters: $\tau_{cycle}$, the mean period between bursts,
$\lambda_{rad,max}$, the maximum value of the dimensionless accretion
rate (at the peak of the burst), and $\langle \lambda_{rad} \rangle$,
the time-averaged dimensionless accretion rate. These parameters are
typically calculated as the mean over $\sim$~5 oscillation cycles and
the error bars represent the standard deviation of the measurements.

After reaching the peak, the luminosity decreases nearly exponentially
on a time scale $\tau_{on}$, that we identify as the duration of the
burst. Both $\tau_{on}$ and the duty cycle, $f_{duty}$, of the black
hole activity ({\it i.e.,} the fraction of time the black hole is
active), can be expressed as a function of $\tau_{cycle}$,
$\lambda_{rad,max}$ and $\langle \lambda_{rad} \rangle$:
\begin{eqnarray}
\tau_{on} &\equiv& \frac{\langle \lambda_{rad}
  \rangle}{\lambda_{rad,max}} \tau_{cycle} , \\
f_{duty} &\equiv& \frac{\tau_{on}}{\tau_{cycle}} = \frac{\langle \lambda_{rad}
  \rangle}{\lambda_{rad,max}}.
\label{eq:fduty} 
\end{eqnarray}
The values of $\lambda_{rad,max}$ and $f_{duty}$ as a function of the
black hole mass, the density and the temperature of the ambient medium
are important for estimating the possibility of detection of IMBHs in
the local universe because these values provide an estimate of the
maximum luminosity and the number of active sources in the local
universe at any time. On the other hand, the mean accretion rate, $\langle \lambda_{rad}
\rangle$ is of critical importance for estimating IMBH growth rate in
the early universe.

\begin{figure*}[thb]
\epsscale{1.1}
\plottwo{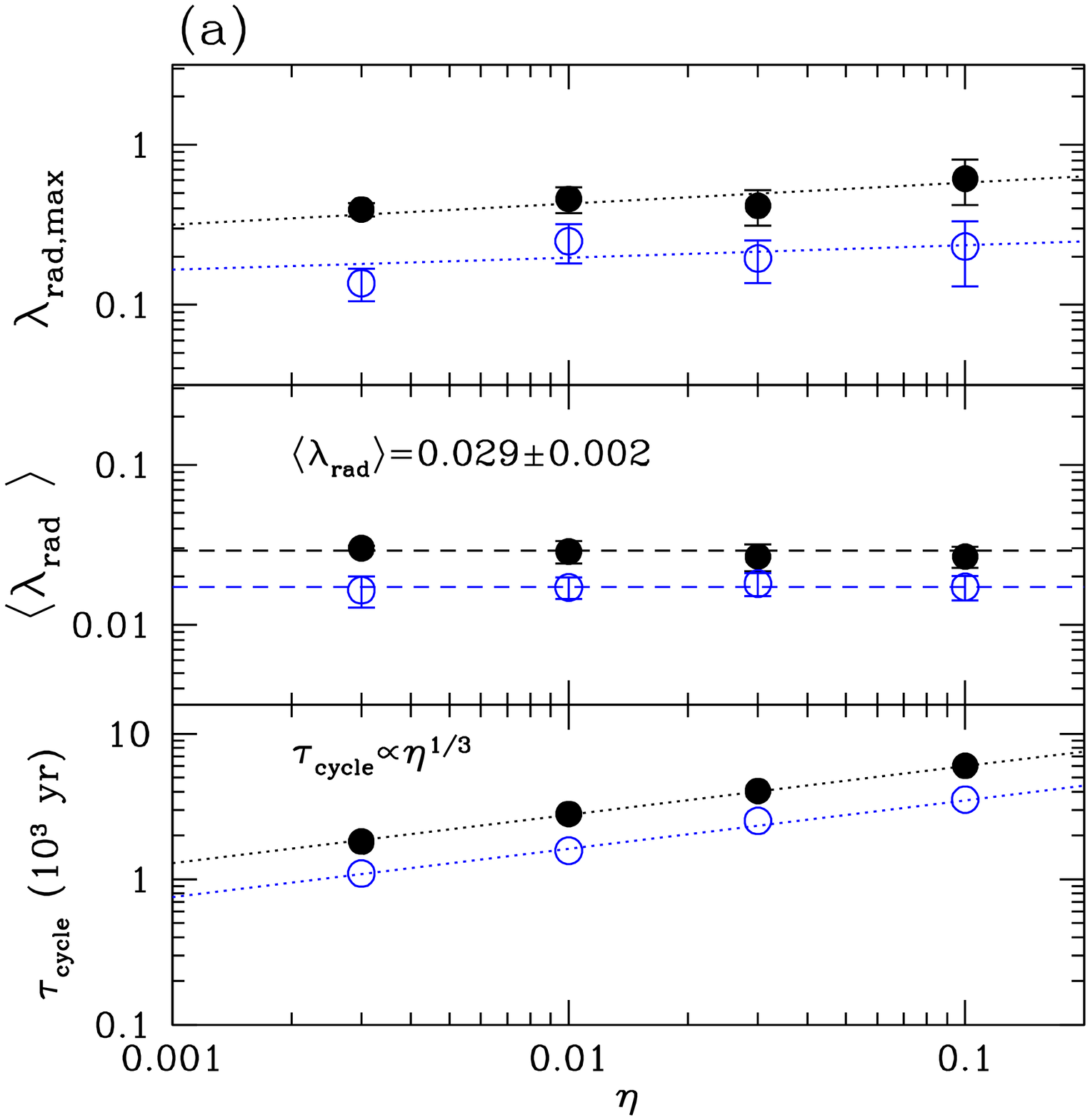}{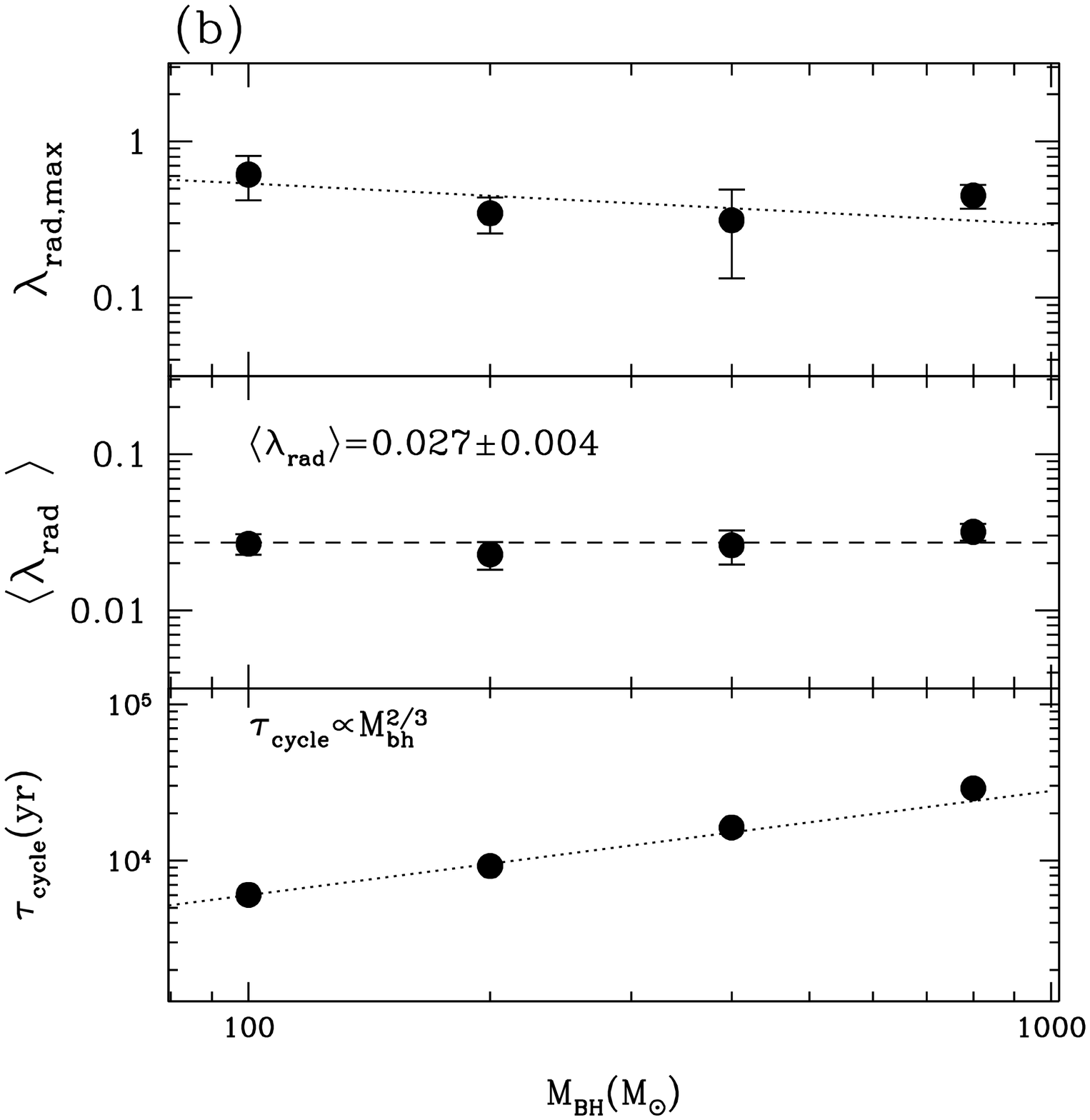}
\plottwo{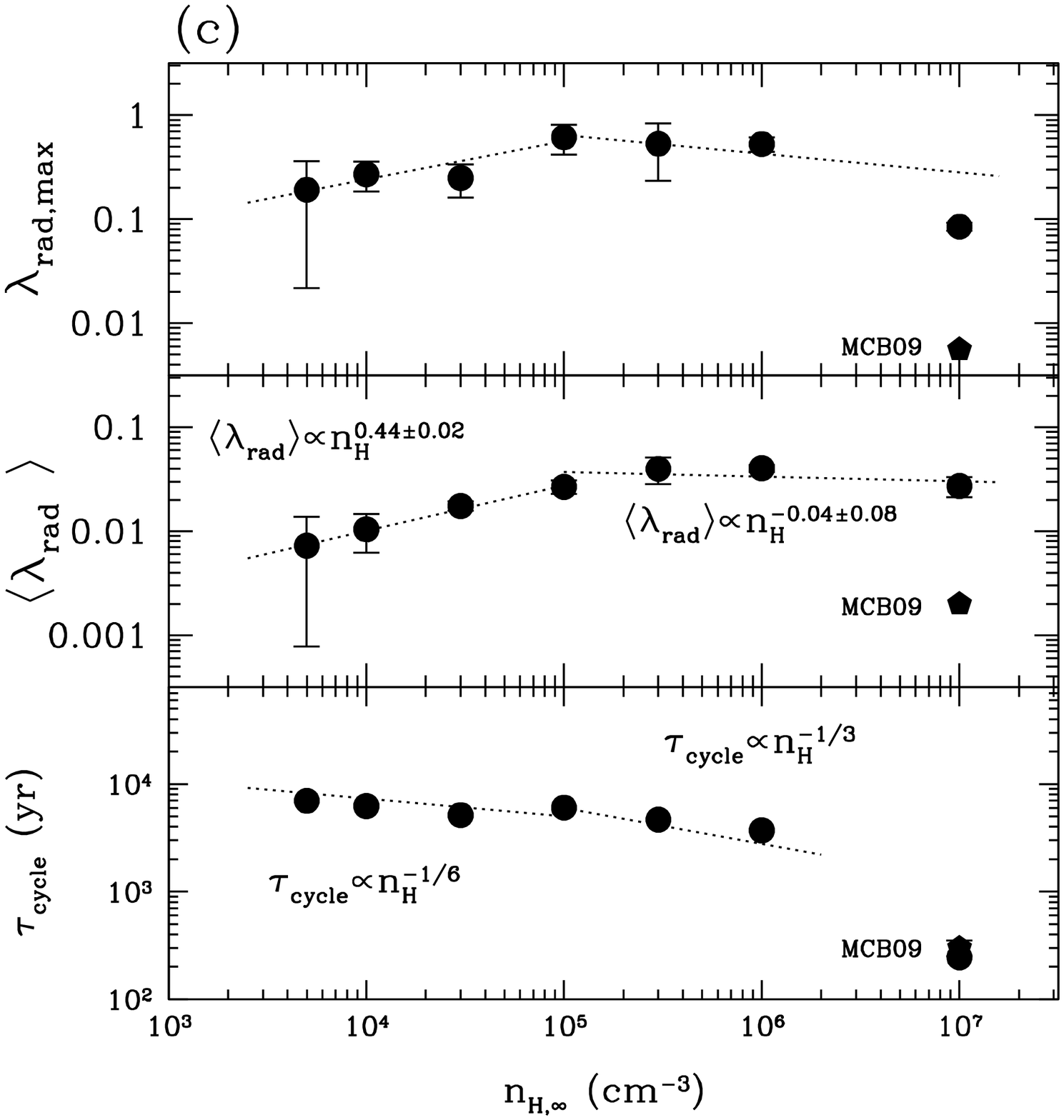}{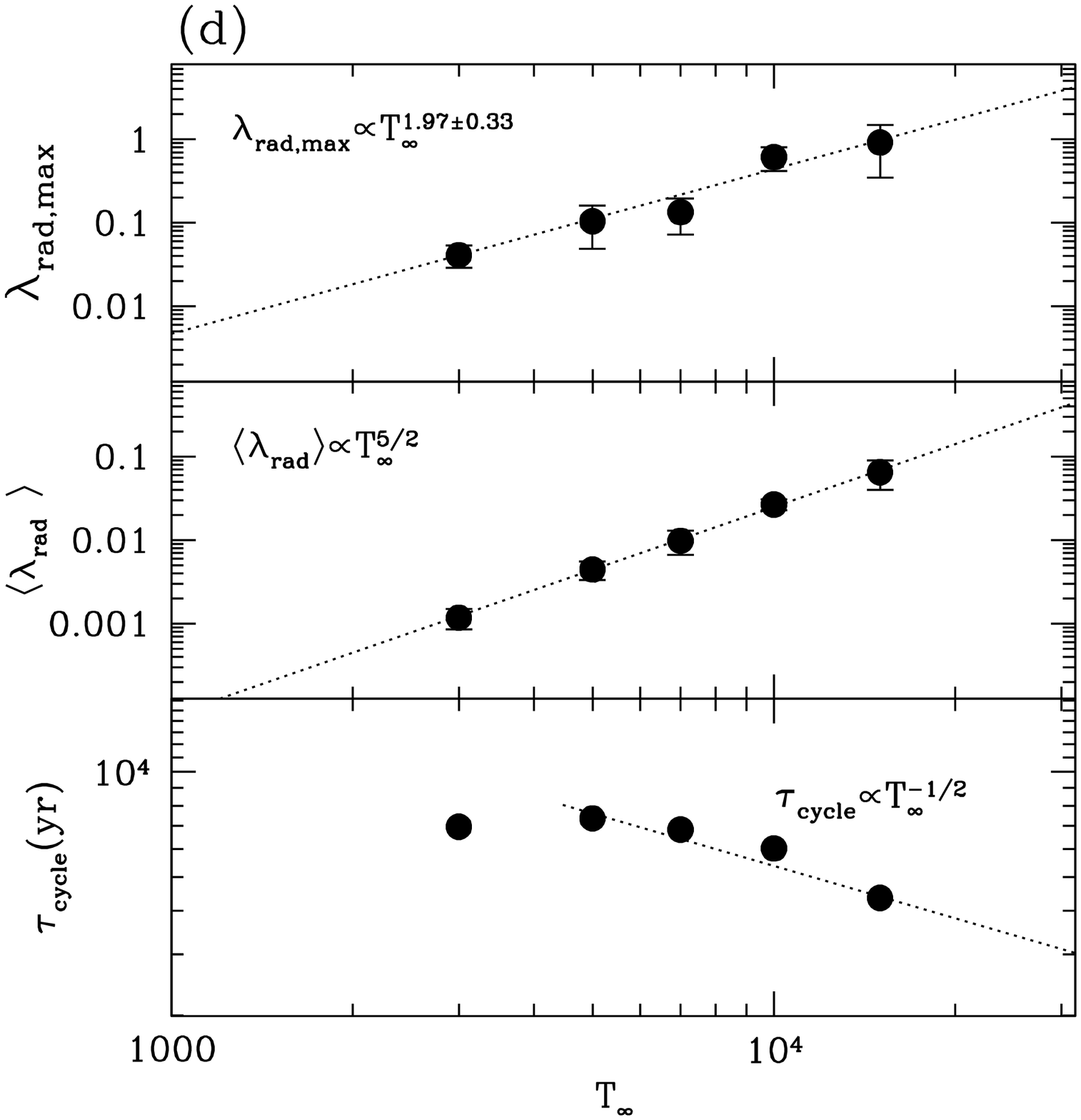}
\caption{ For each panel, peak accretion rate, average accretion rate
  and period between bursts are shown from top to bottom as a function
  of a given parameter. Error bars represent the standard deviation
  around the mean values over $\sim 5$ accretion cycles. {\it (a)}
  Dependence on $\eta$. $\langle \lambda_{rad} \rangle \sim const$
  while $\tau_{cycle} \propto \eta^{1/3}$. Open symbols indicate the
  simulations including helium photo-heating and cooling, which show
  $\sim 41 \% $ lower accretion rate and $ \sim 42 \%$ shorter
  period. {\it (b)} Same plots as a function of $M_{bh}$.  $\langle
  \lambda_{rad} \rangle \sim const$ while $\tau_{cycle} \propto
  M_{bh}^{2/3}$. {\it (c)} Same plots as a function of $n_{H,\infty}$
  of gas. At low densities, $\tau_{cycle} \propto
  n_{H,\infty}^{-1/6}$, whereas at higher density, $\tau_{cycle}
  \propto n_{H,\infty}^{-1/3}$. {\it (d)} Same plots as a function of
  $T_{\infty}$. Average accretion rate $\langle \lambda_{rad} \rangle
  \propto T_{\infty}^{2.5}$. With an exception at lowest temperature
  $\tau_{cycle} \propto T_{\infty}^{-0.5}$.}
\label{para}
\end{figure*}

The four panels in Figure~\ref{para} summarize the results of a set of
simulations in which we vary the free parameters one at a time.  We
find that, in most of the parameter space that we have explored, the
period of the oscillations and the accretion rates are described by a
single or a split power law with slope $\beta$. In the following
paragraphs we report the values of $\beta$ derived from weighted least
squares fitting of the simulation results. The weight is $1/\sigma$
where $\sigma$ is the standard deviation of $\langle \lambda_{rad}
\rangle$ or $\lambda_{rad,max}$ over several oscillations. 

\begin{figure}[thb]
\epsscale{1.0}
\plotone{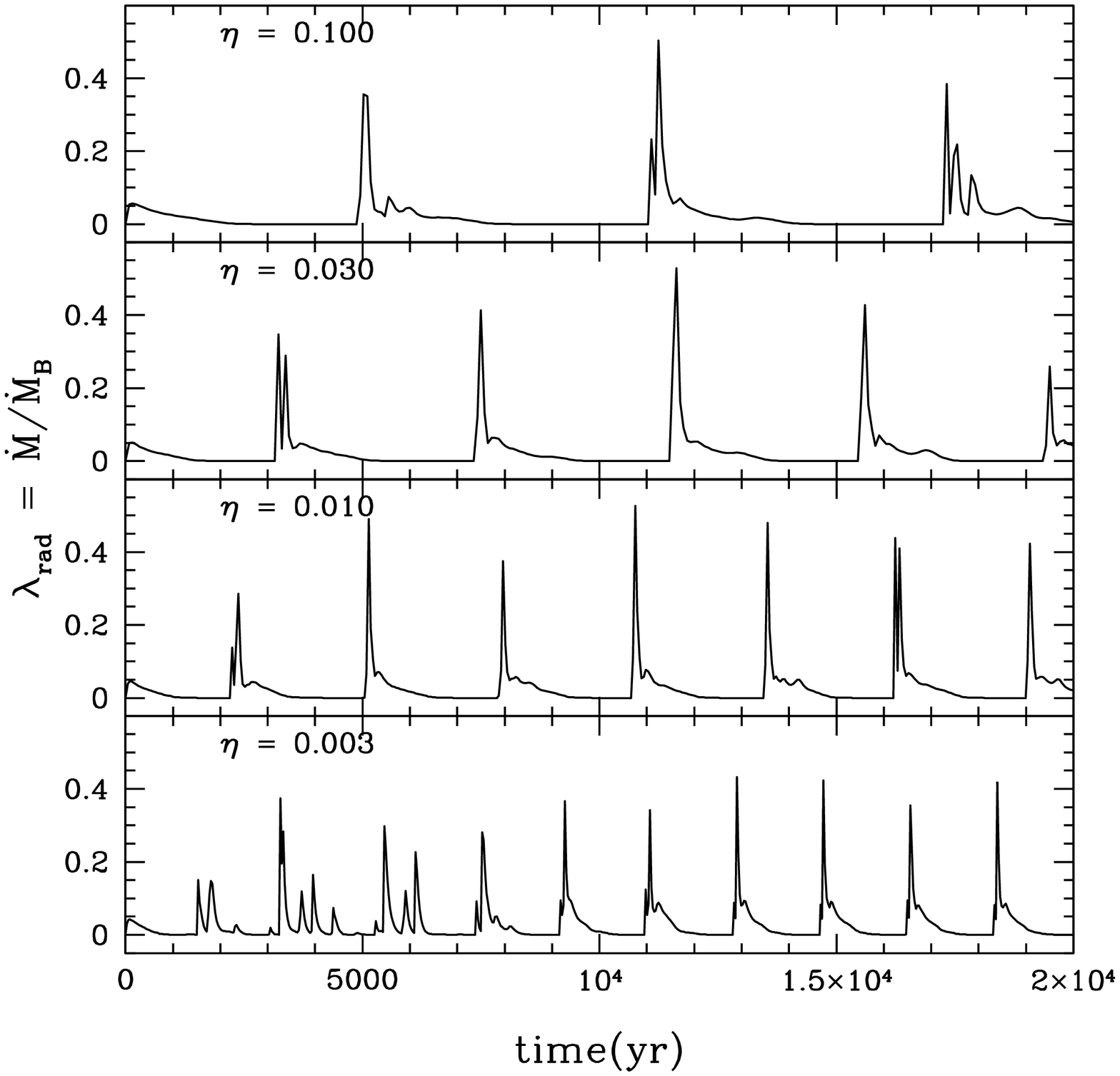}
\caption{Dependence of accretion rate and period of oscillations on
  the radiative efficiency $\eta$. From top to bottom the evolution of
  accretion rate is shown for $\eta=0.1$, $0.03$, $0.01$ and
  $0.003$. The peak accretion rate does not change much with $\eta$,
  but intervals between oscillations decrease with decreasing $\eta$.}
\label{eta1}
\end{figure}

\bigskip\noindent {\it a) Dependence on the radiative efficiency}
\smallskip \\ First, we explore how the accretion depends on the
radiative efficiency $\eta$. This parameter describes the fraction,
$\eta$, of the accreting rest mass energy converted into radiation
while the remaining fraction, $1-\eta$, is added to the black hole
mass. We have explored both constant values of the radiative
efficiency and the case $\eta \propto {\dot m}$ for $l < 0.1$ (see
\S~2.2). The simulation results shown in this section are obtained
assuming $\eta$ is constant. We find similar results for $\langle
\lambda_{rad} \rangle$, $\lambda_{rad,max}$ and $\tau_{cycle}$ when we
assume $\eta \propto \dot{m}$.  The radiative efficiency for a thin
disk is about $10\%$. Here, we vary $\eta$ in the range: $0.2\%$ to
$10\%$. The other free parameters are kept constant with values
$n_{H,\infty}=10^5$~cm$^{-3}$, $M_{bh}=100$~M$_{\sun}$ and
$T_{\infty}=10^4$~K.  Figure~\ref{eta1} shows the accretion rate as a
function of time for different values of the radiative efficiency:
$\eta= 0.1, 0.03, 0.01$ and $0.003$. Panel (a) in Figure~\ref{para}
shows the dependence on $\eta$ of the three parameters that
characterize the accretion cycle. The maximum accretion rate increases
mildly with increasing $\eta$ (log slope $\beta=0.13 \pm 0.06$).  The
average accretion rate is $\langle \lambda_{rad} \rangle \sim 2.9\%
\pm 0.2\% $, is nearly independent of $\eta$ ($\beta
=-0.04\pm0.01$). The period of the oscillations increases with $\eta$
as $\tau_{cycle} \propto \eta^{1/3}$. We also show the simulation
results including helium photo-heating and cooling, shown as open
symbols in the same panel of Figure~\ref{para}. We find that including
helium does not change the qualitative description of the results, but
does offset the mean accretion rate, that is $\sim 41 \%$ lower and
the period of the accretion bursts, that is $ \sim 42\%$ shorter. This
offset of the accretion rate and period with respect to the case
without helium is due to the higher temperature of the gas inside the
HII region surrounding the black hole.

\bigskip\noindent
{\it b) Dependence on black hole mass}
\smallskip
\\ We explore a range in black hole mass from 100~M$_{\sun}$ to
800~M$_{\sun}$, while keeping the other parameters constant ($\eta =
0.1$, $n_{H,\infty}=10^5$~cm$^{-3}$ and $T_{\infty}= 10^4$~K). The
results are shown in panel (b) of Figure~\ref{para}. The mean
accretion rate is $\langle \lambda_{rad} \rangle \sim 2.7\% \pm 0.4\%$
and the maximum accretion rate is $\lambda_{rad,max} \sim 42\% \pm
12\% $ ($\beta = -0.26 \pm 0.20$). They are both independent of
$M_{bh}$ within the error of the fit. The period of the bursts is well
described by a power-law relation $\tau_{cycle} \propto M_{bh}^{2/3}$.


\bigskip\noindent 
{\it c) Dependence on gas density of the ambient
  medium} 
\smallskip 
\\ Panel (c) in Figure~\ref{para} shows the dependence of accretion
  rate and burst period on the ambient gas density, $n_{H,\infty}$.
  We explore a range of $n_{H,\infty}$ from $5\times 10^3$~cm$^{-3}$
  to $10^7$~cm$^{-3}$, while keeping the other parameters constant at
  $\eta = 0.1$, $M_{bh}=100$~M$_{\sun}$ and $T_{\infty}= 10^4$~K. For
  densities $n_{H,\infty} \ge 10^5$~cm$^{-3}$, $\langle \lambda_{rad}
  \rangle$ and $\lambda_{rad, max}$ are insensitive to $n_{H,\infty}$
  ($\beta = -0.04\pm0.08$ and $\beta=-0.18\pm0.13$, respectively).
  However, for $n_{H,\infty} \le 10^5$~cm$^{-3}$, $\langle
  \lambda_{rad}\rangle$ and $\lambda_{rad, max}$ are proportional to
  $n_{H,\infty}^{1/2}$ ($\beta = 0.44 \pm 0.02$ and $\beta = 0.37 \pm
  0.09$, respectively). 


The bottom of Figure~\ref{para}(c) shows the effect of density in
determining the oscillation period. For densities $n_{H,\infty} \ge
10^5$~cm$^{-3}$, $\tau_{cycle}$ is fitted well by a power law with
$\tau_{cycle} \propto n_{H,\infty}^{-1/3}$ and for the densities
$n_{H,\infty} \le 10^5$~cm$^{-3}$ it is fitted well by a power law
$\tau_{cycle} \propto n_{H,\infty}^{-1/6}$. However, $\tau_{cycle}$ at
$n_{H,\infty}=10^7~$cm$^{-3}$ is lower than predicted by the power law
fit for $n_{H,\infty} \ge 10^5~$cm$^{-3}$. Although Figure \ref{1d2d},
\ref{eta1} do not show clearly the magnitude of accretion rate during
the inactive phase, it is evident in a log-log plot that accretion
rate at minima is 4 orders of magnitude lower than during the peak of
the burst. This is the case for all simulations but the ones with
$n_{H,\infty}=10^7~$cm$^{-3}$ in which the accretion rate at minima is
2 orders of magnitude higher than in all other simulations.  The
simulations show that the ambient gas density is an important
parameter in determining the accretion luminosity and period between
bursts of the IMBH. One of the reasons is that the gas temperature
inside the hot ionized bubble and the thickness and density of the
dense shell in front of it depend on the density via the cooling
function. The drop in the accretion rate we observe at low densities
can be linked to an increase of the temperature within the sonic
radius with respect to simulations with higher ambient density. This
results in an increase in the pressure gradient within the ionized
bubble that reduces the accretion rate significantly.

\bigskip\noindent
{\it d) Dependence on the temperature of the ambient medium}
\smallskip
\\ Panel (d) in Figure~\ref{para} shows the dependence of accretion
rate and period of the bursts on the temperature of the ambient
medium, $T_\infty$.  We vary $T_{\infty}$ from $3000$~K to $15000$~K
while keeping the other parameters constant at $\eta = 0.1$,
$M_{bh}=100$~M$_{\sun}$ and $n_{H,\infty}= 10^5$~cm$^{-3}$.  We find
$\langle \lambda_{rad} \rangle$ and $\lambda_{rad,max}$ depend steeply
on $T_\infty$ as $T_{\infty}^{5/2}$ ($\beta=2.44 \pm 0.06$). Except
for the simulation with $T_\infty=3000$~K, the period of the accretion
cycle is fitted well by a single power law $\tau_{cycle}\propto
T_{\infty}^{-1/2}$.

\section{Analytical Formulation of Bondi Accretion with Radiative Feedback}

In this section we use the fitting formulas for $\langle \lambda_{rad}
\rangle$, $\lambda_{rad, max}$ and $\tau_{cycle}$ obtained from the
simulations, to formulate an analytic description of the accretion
process.  For ambient densities $n_{H,\infty} \ge 10^5$~cm$^{-3}$, we
have found that the dimensionless mean accretion rate $\langle
\lambda_{rad} \rangle$ depends only on the temperature of the ambient
medium. It is insensitive to $\eta$, $M_{bh}$ and
$n_{H,\infty}$. Thus, for $n_{H,\infty} \ge10^5$~cm$^{-3}$ we find
\begin{eqnarray}
  \langle \lambda_{rad} \rangle \sim
  3.3\%~T_{\infty,4}^{5/2}~n_{H,5}^{-0.04} \sim
  3.3\%~T_{\infty,4}^{5/2} ,
\label{eq:alambda_a}
\end{eqnarray}
 while for $n_{H,\infty} \le 10^5$~cm$^{-3}$ we find
\begin{eqnarray}
  \langle \lambda_{rad} \rangle \sim 3.3\%~
  T_{\infty,4}^{5/2}~n_{H,5}^{1/2}.
\label{eq:alambda_b}
\end{eqnarray}

As mentioned above, the dependence of $\langle \lambda_{rad} \rangle$
on the density is due to the increasing temperature inside the ionized
bubble at low densities.
The period of the accretion cycle depends on all the
parameters we have investigated in our simulation. In the range of
densities $n_{H,\infty} \ge 10^5$~cm$^{-3}$ we find
\begin{eqnarray}
\tau_{cycle} &=& (6\times 10^3~{\rm yr})~\eta_{-1}^{\frac{1}{3}}~
M_{bh,2}^{\frac{2}{3}}~n_{H,5}^{-\frac{1}{3}}~T_{\infty,4}^{-\frac{1}{2}}
\end{eqnarray}
where we use the notation of
$M_{bh,2}\equiv M_{bh}/(10^2$~M$_{\sun})$. However, at lower densities
$n_{H,\infty} \le 10^5$~cm$^{-3}$, we find
\begin{eqnarray}
\tau_{cycle} &=& (6\times 10^3~{\rm yr})~\eta_{-1}^{\frac{1}{3}}~
M_{bh,2}^{\frac{2}{3}}~n_{H,5}^{-\frac{1}{6}}~T_{\infty,4}^{-\frac{1}{2}} 
\end{eqnarray}
in which only the dependence on $n_{H,5}$ changes. The different
dependence of $\tau_{cycle}$ on $n_{H,\infty}$ is associated with a
change of the mean accretion rate $\langle \lambda_{rad} \rangle$ for
each density regime. The deviation of $\tau_{cycle}$ from the power
law fit at $n_{H,\infty} =10^7$~cm$^{-3}$ is not associated with any
variation of the mean accretion rate. Our value of $\tau_{cycle}$ for
$n_{H,\infty} =10^7$~cm$^{-3}$ is in good agreement with the value
found by MCB09.


\subsection{ Dimensionless accretion rate : $\langle \lambda_{rad}
  \rangle $}

In this section we seek a physical explanation for the relationship
between the mean accretion rate $\langle \lambda_{rad} \rangle$ and
the temperature of the ambient medium found in the simulations. The
model is valid in all the parameter space we have explored with a
caveat in the low density regime ($n_{H,\infty}<3\times
10^5$~cm$^{-3}$) and at low ambient temperatures
($T_{\infty}<3000$~K).

\begin{figure*}[thb]
\epsscale{1.0}
\plotone{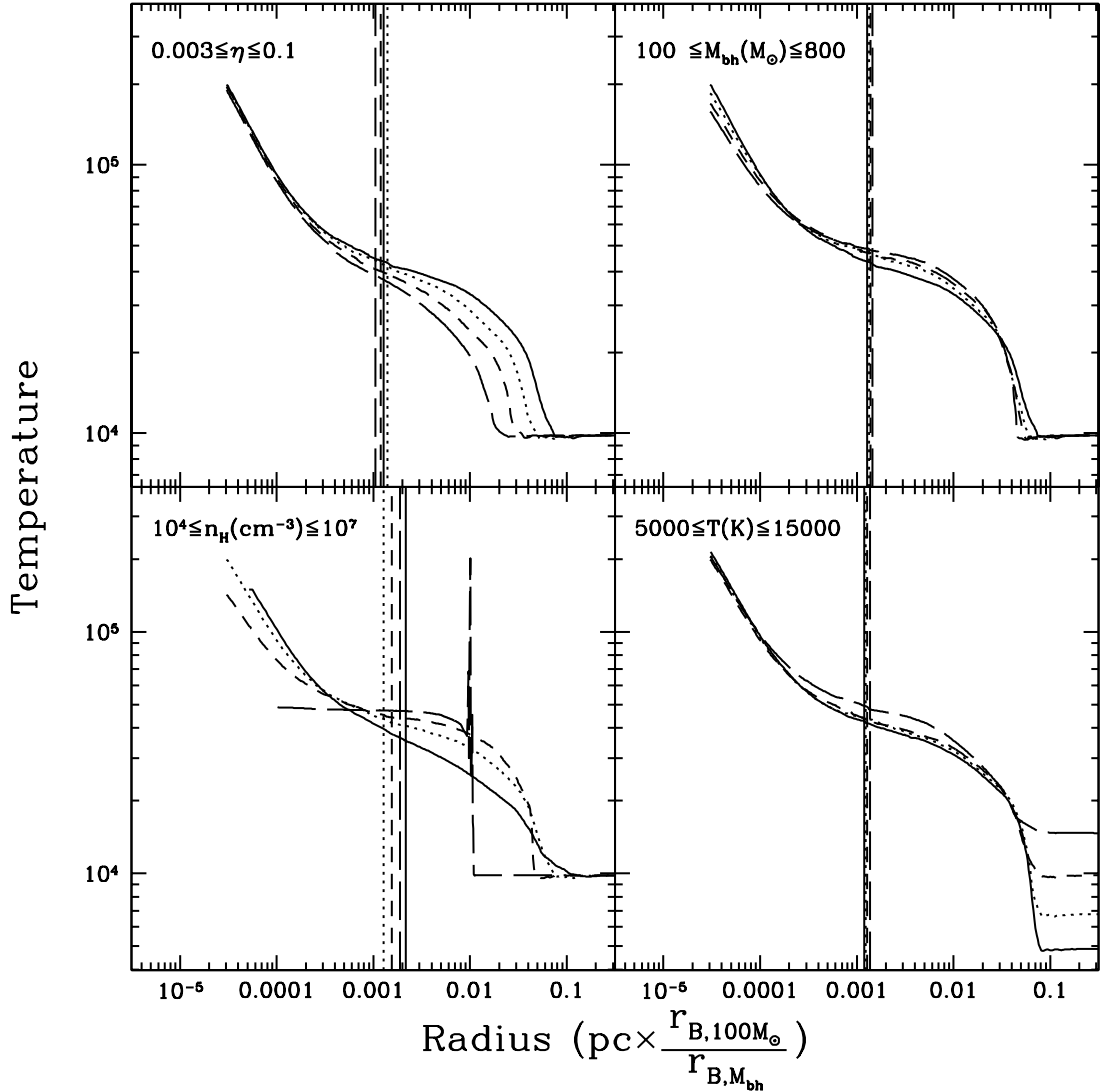}
\caption{ Time-averaged temperature profiles as a function of
  simulation parameters. Different vertical lines indicate accretion
  radii, $r_{acc}$, for each parameter. {\it Top left }: $ \eta $
  ranges from 0.1 ({\it solid line}) to 0.003 ({\it long
  dashed line}).  {\it Top right }: $M_{bh}$ ranges from 100
  M$_{\sun}$ ({\it solid}) to 800 M$_{\sun}$ ({\it long dashed}). {\it
  Bottom left }: Density ranges from $10^4$~cm$^{-3}$ ({\it solid}) to
  $10^7$~cm$^{-3}$ ({\it long dashed}).  {\it Bottom right }:
  $T_{\infty}$ ranges from $5000$K ({\it solid}) to $15000$K ({\it
  long dashed}).  }
\label{temp_all}
\end{figure*}

Figure~\ref{temp_all} shows the time-averaged temperature profiles for
simulations in which we vary $\eta$, $M_{bh}$, $n_{H,\infty}$ and
$T_{\infty}$. In the case of different $M_{bh}$ the radii are rescaled
so that direct comparisons can be made with the case of
$100~$M$_{\sun}$. Vertical lines indicate the accretion radius
$r_{acc}$, inside of which gas is accreted and outside of which gas is
pushed out to the ionization front. We find that the value of
$r_{acc}$ is generally insensitive to the parameters of the
simulation as is the gas temperature at $r_{acc}$. 

Accretion onto the black hole of gas inside the hot ionized sphere is
limited by the thermal pressure of the hot gas and by the outflow
velocity of the gas that is produced by the pressure gradient inside
the Str\"{o}mgren sphere. Thus, the accretion radius, $r_{acc}$, is
analogous to the inner Bondi radius, $r_{b,in}$, modified to take into
account temperature and pressure gradient inside the hot bubble. 

Let us assume that the average accretion rate onto the black hole is 
\begin{equation}
\langle {\dot M} \rangle = 4\pi \lambda_{B}
r_{acc}^2\rho_{in}c_{s,in},
\label{eq:acc}
\end{equation}
where $\rho_{in}$ and $c_{s,in}$ (and the corresponding temperature
$T_{in}$) are the density and the sound speed at $r_{acc}$. Based on
the results illustrated in Figure~\ref{temp_all}, we expect the
accretion rate to depend only on $\rho_{in}$, since $r_{acc}$ and
$c_{s,in}$ can be taken to be constants.
 
When a Str\"{o}mgren sphere is formed, the gas inside the hot bubble
expands and its density decreases. Inside the ionization front the
temperature is about $10^4-10^5$K. Thus, assuming pressure equilibrium
across the ionization front we find the dependence of $\rho_{in}$ on
$T_{\infty}$:
\begin{eqnarray}
  \rho_{in} \approx \rho_{\infty}{T_{\infty} \over T_{in}}=\rho_\infty
  \left({c_{s,\infty} \over c_{s,in}}\right)^{2}. 
\label{eq:pres_eq}
\end{eqnarray}
We find $f=r_{acc}/r_{b,in} \sim 1.8$ and the temperature at $r_{acc}$
is $T_{in}\sim~4\times 10^4$~K independent of $\eta$, $M_{bh}$,
$n_{H,\infty}$ and $T_{\infty}$ for a fixed spectral index of
radiation $\alpha=1.5$. The dimensionless accretion rate inside of the
Str\"{o}mgren sphere normalized by the Bondi accretion rate in the
ambient medium is then :
\begin{eqnarray}
\langle \lambda_{rad} \rangle &\simeq& \lambda_B \frac{r_{acc}^2
    \rho_{in} c_{s,in}}{r_{b,\infty}^2\rho_{\infty}c_{s,\infty}} \nonumber \\
    &\simeq& \frac{1}{4} (1.8)^2\left( \frac{\rho_{in}}{\rho_{\infty}}
    \right) \left( \frac{c_{s,in}}{c_{s,\infty}} \right)^{-3} \nonumber \\
    &\simeq& 3\%~T_{\infty,4}^{2.5} 
\label{eq:alambda}
\end{eqnarray}
where we have used $\lambda_{B}=1/4$ appropriate for an adiabatic gas.
Thus, $\langle \lambda_{rad} \rangle \propto T_{\infty}^{5/2}$ which
is in agreement with the simulation result, given that $r_{acc}$ and
$T_{in}$ remain constant when we change the simulation
parameters. However, $r_{acc}$ and $T_{in}$ may not stay constant if
we modify the cooling or heating function, for instance by increasing
the gas metallicity or by changing the spectrum of radiation; this
result suggests that the accretion rate is very sensitive to the
details of the temperature structure inside the Str\"{o}mgren sphere
which shows a dependence on $n_{H,\infty}$. The temperature profile
changes significantly for $n_{H,\infty}<3\times 10^4$~cm$^{-3}$ and
this is probably the reason why our model does not fit perfectly
$\langle \lambda_{rad} \rangle$ from the simulations in the lower
density regime. In the next section we test whether
Equation~(\ref{eq:acc}) is still a good description of our results
when we change the thermal structure inside the HII region.

\subsubsection{Dependence on temperature at accretion radius}

\begin{figure*}[thb]
\epsscale{1.01} \plottwo{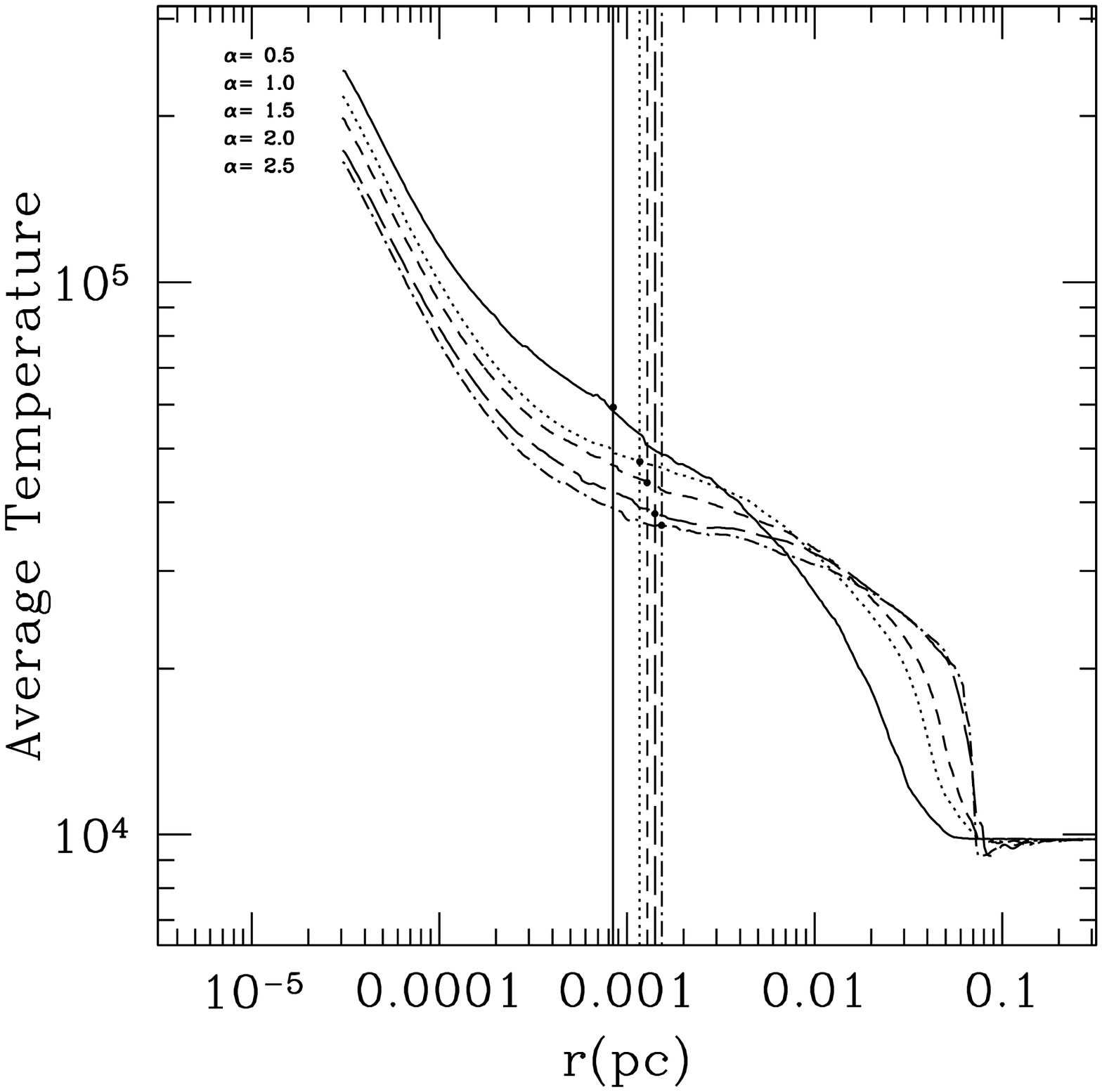}{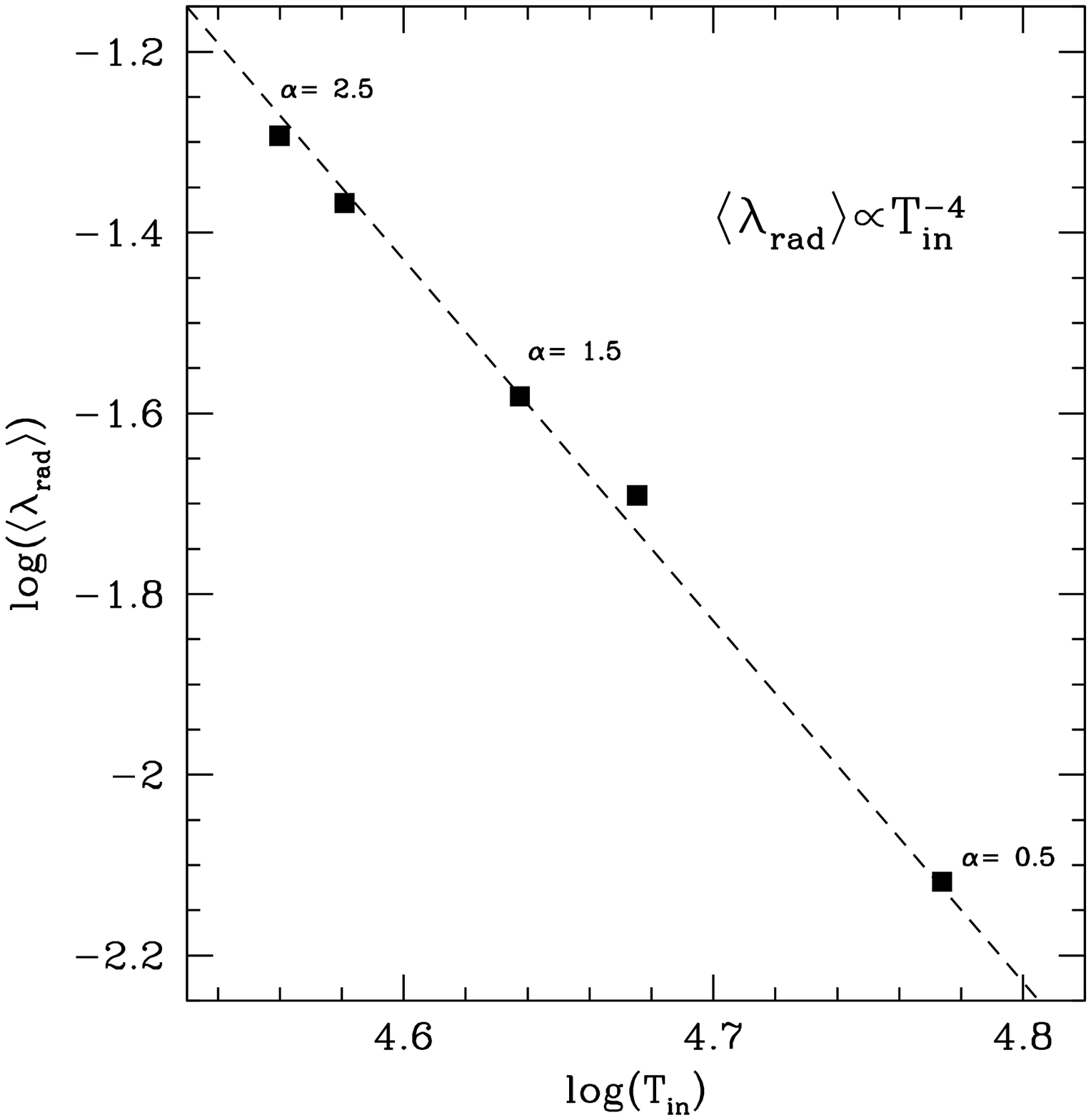}
\caption{ {\it Left} : Average temperature profiles of the HII region
  as a function of spectral index $\alpha$. Smaller $\alpha$ results
  in smaller $r_{acc}$ and higher $T_{in}$. {\it Right} : Relation
  between temperature at $r_{acc}$ and average accretion rate $\langle
  \lambda_{rad} \rangle $. We find $\langle \lambda_{rad} \rangle
  \propto T_{in}^{-4}$.}
\label{pr10_alpha}
\end{figure*}

In this section we study the dependence of the accretion rate on the
time-averaged temperature $T_{in}$ at $r_{acc}$. We change the
temperature $T_{in}$ by varying the spectral index $\alpha$ of the
radiation spectrum and by including Compton cooling of the ionized gas
by CMB photons. Here we explore the spectral index of the radiation
spectrum in the range $\alpha = 0.5,1.0,1.5,2.0,2.5$ with the energy
of photons from $10$~keV $100$~keV. The other parameters are kept
constant at $\eta = 0.1$, $M_{bh}=100$~M$_{\sun}$,
$n_{H,\infty}=10^{5}$~cm$^{-3}$ and $T_{\infty}= 10^4$~K.

Figure~\ref{pr10_alpha} shows the different time-averaged temperature
profiles for different values of $\alpha$. Spectra with lower values
of the spectral index $\alpha$ produce more energetic photons for a
given bolometric luminosity, increasing the temperature inside the
ionized bubble. Simulations show that the averaged accretion rate
$\langle \lambda_{rad} \rangle$ increases for softer spectrum of
radiation. Different slopes ($0.5 \le \alpha \le 2.5$) of the
power-law spectrum lead to different $T_{in}$ ($59000~$K to
$36000~$K) and $\langle \lambda_{rad} \rangle$ (0.0076 to 0.0509). 
Adopting a harder spectrum (with $\alpha = 0.5$) instead of the softer
($\alpha = 2.5$) increases $T_{in}$ 
by a factor of $1.6$ and $\langle \lambda_{rad} \rangle$ decreases by
a factor $6.7$. The fit to the simulation results in
Figure~\ref{pr10_alpha} show that $\langle \lambda_{rad} \rangle$
depends on temperature at $r_{acc}$ as
\begin{eqnarray}
\langle \lambda_{rad} \rangle  \propto T_{in}^{-4} \propto c_{s,in}^{-8}.
\end{eqnarray}

The dependence on $c_{s,in}$ differs from equation (\ref{eq:alambda}).
However this is not surprising because in these simulations the values
of $r_{acc}$ and $c_{s,in}$ do not remain constant while we vary the
value of the spectral index $\alpha$. This is due to a change of the
temperature and pressure gradients within the HII region. The
accretion radius, $r_{acc}$, can be expressed as a function of the
Bondi radius inside the hot bubble,
$r_{b,in}=GM_{bh}c_{s,in}^{-2}$. From the simulations we obtain the
following relationship between these two radii: 
\begin{eqnarray}
f=\frac{r_{acc}}{r_{b,in}} \simeq 1.8
\left(\frac{T_{in}}{4\times10^4~{\rm K}} \right)^{-0.7\pm0.2}.
\end{eqnarray}

Thus, if our model for the
accretion rate summarized by equation~(\ref{eq:acc}) is valid, we
should have:
\begin{eqnarray}
  \langle \lambda_{rad} \rangle &\simeq& \frac{1}{4} \frac{r_{acc}^2
    \rho_{in} c_{s,in}}{r_{b,\infty}^2\rho_{\infty}c_{s,\infty}}
  \simeq \frac{(1.8)^2}{4}\left( \frac{\rho_{in}}{\rho_{\infty}}
  \right) c_{s,in}^{-5.9} c_{s,\infty}^3 \nonumber \\ &\simeq&
3\%~T_{\infty,4}^{2.5} \left(\frac{T_{in}}{4\times10^4~{\rm K}}\right)^{-4},
\end{eqnarray}
in agreement with the simulation results $\langle \lambda_{rad}
\rangle \propto T_{\infty}^{2.5} T_{in}^{-4}$ where the dependence on
$T_{in}$ was not explored initially. Thus, Bondi-like accretion on the
scale of $r_{acc}$ is indeed a good explanation of our results. Given
the steep dependence of the value of accretion rate $\langle
\lambda_{rad} \rangle$ on $T_{in}$ it is clear that it is very
sensitive on the details of the thermal structure inside the HII
region. This means that $\langle \lambda_{rad}\rangle$ depends on the
spectrum of radiation and gas metallicity.

\subsection{Accretion rate at peaks and duty cycle: $\lambda_{rad,max}$, $f_{duty}$}
We estimate $f_{duty}$ by comparing $\lambda_{rad,max}$ and $\langle
\lambda_{rad} \rangle$ using Equation (\ref{eq:fduty}). This quantity
gives an estimate of what fraction of black holes are accreting gas at
the rate close to the maximum. Within the fitting errors, the log
slopes of $\lambda_{rad,max}$ and $\langle \lambda_{rad} \rangle$ as a
function of the parameters $M_{bh}$, $T_{\infty}$ are zero. Thus, we
assume that the dimensionless accretion rates are independent of these
parameters.

For $n_{H,\infty} \ge 10^5$~cm$^{-3}$, $\lambda_{rad,max}$ can be
expressed as $\lambda_{rad,max} \sim
0.55~\eta_{-1}^{0.13}~n_{H,5}^{-0.18}~T_{\infty, 4}^{2.0} $ and the
dependence of $f_{duty}$ on these parameters can be expressed using
equation (\ref{eq:alambda_a}) as
\begin{eqnarray}
f_{duty} &\sim& 6\%~\eta_{-1}^{-0.13}~n_{H,5}^{0.14}~T_{\infty,4}^{0.5}
\end{eqnarray}
where we include the mild dependence of $\langle \lambda_{rad}
\rangle$ on the density. For $n_{H,\infty}\le~10^5$~cm$^{-3}$,
$\lambda_{rad,max} \sim
0.55~\eta_{-1}^{0.13}~n_{H,5}^{0.37}~T_{\infty,4}^{2.0}$ has a
different power law dependence on the density and we get $f_{duty}$ as
\begin{eqnarray}
f_{duty} &\sim& 6\%~\eta_{-1}^{-0.13}~n_{H,5}^{0.07}~T_{\infty,4}^{0.5}
\end{eqnarray}
where $f_{duty}$ shows a milder dependence on the gas density.  Thus,
we expect about 6\% of IMBHs to be accreting near the maximum rate at
any given time. This value depends weakly on $\eta$, $n_{H,\infty}$
and $T_{\infty}$.

\subsection{ Average period between bursts : $\tau_{cycle}$ }

\begin{figure}[thb]
\epsscale{1.0} 
\plotone{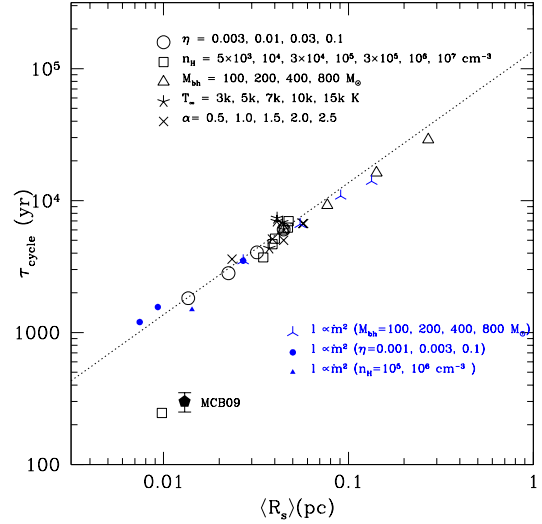}
\caption{Period of accretion bursts as a function of the average
  Str\"{o}mgren radius. All simulation results from all the parameters
  are plotted together.  The average size of the Str\"{o}mgren sphere
  shows a linear relation with the period $\tau_{cycle}$. The only
  exception happens at the highest density
  ($n_{H,\infty}=10^7$~cm$^{-3}$), but this result is in agreement
  with the work by MCB09 (symbol with error bar)}
\label{rs_period}
\end{figure}

In this section we derive an analytical expression for the period of
the luminosity bursts as a function of all the parameters we
tested. Although $\tau_{cycle}$ shows a seemingly complicated power
law dependencies on the free parameters, we find that $\tau_{cycle}$
is proportional to the time-averaged size of the Str\"{o}mgren
sphere. This is shown in Figure~\ref{rs_period}. The linear relation
between $\tau_{cycle}$ and the average Str\"{o}mgren radius $\langle
R_s \rangle$ explains the dependence of $\tau_{cycle}$ on every
parameter considered in this work.

The number of ionizing photons created by accretion onto a black hole
is determined by the average accretion rate and the radiative
efficiency $\eta$. The average accretion rate itself can be expressed
as a fraction of the Bondi accretion rate $\langle \lambda_{rad}
\rangle$. Therefore, the average number of ionizing photons emitted near the
black hole can be expressed as
\begin{eqnarray}
N_{ion} &\propto& \eta \langle \lambda_{rad} \rangle \dot{M}_{B} 
\\ \nonumber &\propto& \eta \langle \lambda_{rad} \rangle \frac{G^2
  M_{bh}^2 }{c_{s,\infty}^3} \rho_{\infty} .
\end{eqnarray}
It follows that:
\begin{eqnarray}
\tau_{cycle} &=& t_{out} \approx {\langle R_s \rangle \over v_{out}}
\propto \left( \frac{3 N_{ion}}{4\pi\alpha_{rec}
n_H^2}\right)^{\frac{1}{3}} \nonumber \\ &\propto& \left(
\frac{1}{n_H^2}\right)^{\frac{1}{3}} \left( \eta \langle \lambda_{rad}
\rangle \frac{G^2 M_{bh}^2 }{c_{s,\infty}^3} \rho_{\infty}
\right)^{\frac{1}{3}},
\end{eqnarray}
where we find $v_{out} \sim {1 \over 3} c_{s,in}$. Ignoring constant
coefficients and using equation~(\ref{eq:alambda_a}) for $n_{H,\infty} \ge
10^5$~cm$^{-3}$, we find :
\begin{eqnarray}
\tau_{cycle} &\propto& \eta^{\frac{1}{3}} M_{bh}^{\frac{2}{3}}
n_{H,\infty}^{-\frac{1}{3}} T_{\infty}^{-\frac{1}{2}},  
\end{eqnarray}
or using equation~(\ref{eq:alambda_b}) for $n_{H,\infty} \le
10^5$~cm$^{-3}$, we find :
\begin{eqnarray}
\tau_{cycle} &\propto& \eta^{\frac{1}{3}} M_{bh}^{\frac{2}{3}}
n_{H,\infty}^{-\frac{1}{6}} T_{\infty}^{-\frac{1}{2}} 
\end{eqnarray} 
which are exactly as in the empirical fitting formulas in both density
regimes and also in good agreement with the analytical work by
MBCO09. This explains the dependence of $\tau_{cycle}$ on any tested
parameter $\eta$, $M_{bh}$, $n_{H,\infty}$ and $T_{\infty}$. In
Figure~\ref{rs_period} we also show simulation results assuming $\eta
\propto \dot{m}$. All simulations show the same relationship between
$\tau_{cycle}$ and $\langle R_s\rangle$. However, the simulation with
the highest ambient density ($n_{H,\infty}=10^7$~cm$^{-3}$) deviates
from the linear relationship, but is in agreement with the numerical
simulation by MCB09. It appears that in the high density regime
$\tau_{cycle}$ decreases steeply with decreasing $\langle R_s\rangle$.

We can interpret $\tau_{cycle}$ as the time scale at which the gas
inside HII region gets depleted. If the gas depletion inside the
Str\"{o}mgren sphere is dominated by the outward gas flow, then
$\tau_{cycle} \propto \langle R_{s} \rangle/c_{s,in} $ in agreement
with the empirical linear relation in Figure \ref{rs_period}. However,
the depletion time scale may be different if the accretion by the
black hole dominates gas consumption inside the Str\"{o}mgren
sphere. We can derive this time scale as
\[
t_{in}= {M_{HII} \over \dot{M}}= \left({\langle R_s \rangle \over
  r_{acc}}\right)^2 {\langle R_s \rangle \over 3~c_{s,in}} \sim
\left({\langle R_s \rangle \over r_{acc}}\right)^2 {t_{out} \over 9}.
\]
Roughly, we expect $\tau_{cycle}=\min{(t_{out}, t_{in})}$. So, for
$\langle R_s \rangle/ r_{acc} \le 3$, the period of the cycle scales
as $\langle R_s \rangle^3$. This may explain the deviation of the
period for $n_{H,\infty}=10^7$~cm$^{-3}$ from the linear relation. We
see in Figure~\ref{temp_all} that the ratio $\langle R_s \rangle
/r_{acc} \sim 5$ for $n_{H,\infty}=10^7$~cm$^{-3}$ which is much
smaller than the ratio found for other densities.

\subsubsection{Rayleigh-Taylor instability}
In 2D simulations we find that RT instability develops across the
Str\"{o}mgren radius, but it decays on short time scales. This can be
explained by the pressure gradient inside the Str\"{o}mgren sphere
which does not allow the RT grow. 
In the linear regime the growth time scale of the RT instability of
wavelength $\lambda$ is 
\[
\tau_{RT} \simeq \sqrt{ {{\rho_{sh}+\rho_{in}} \over
  {\rho_{sh}-\rho_{in}}}
\frac{2\pi \lambda}{g}} \simeq \sqrt{\frac{2\pi \lambda}{g}}
\]
where $\rho_{sh}$ is the density of the shell and $g \simeq
GM_{bh}\langle R_s \rangle ^{-2}$ is the gravitational acceleration at
the shell radius. Thus, RT timescale can be expressed as
\begin{eqnarray}
\tau_{RT} 
\simeq \frac{ \langle R_s \rangle}{c_{s,in}} \sqrt{\frac{2\pi\lambda}{r_{b,in}}}.
\end{eqnarray}
So during one cycle perturbations grow on scales:
\[
\lambda_{RT} < \left( {\tau_{RT} \over \tau_{cycle}} \right) ^2
\frac{r_{b,in}}{2\pi} < \frac{r_{b,in}}{2\pi} 
\] 
where $r_{b,in}$ is the inner Bondi radius. Thus only instability on
angular scales $\theta \sim \lambda_{RT}/2\pi \langle R_s
  \rangle \le r_{b,in}/(2\pi)^2 \langle R_s \rangle$ grow in
our simulation.

\section{Summary and Discussion}
In this paper we simulate accretion onto IMBHs regulated by radiative
feedback assuming spherical symmetric initial conditions. We study
accretion rates and feedback loop periods while varying radiative
efficiency, mass of black hole, density and temperature of the medium,
and spectrum of radiation. The aim of this work is to simulate
feedback-regulated accretion in a wide range of the parameter space to
formulate an analytical description of processes that dominate the
self-regulation mechanism. Thus, in this first paper we keep the
physics as simple as possible, neglecting the effect of angular
momentum of the gas, radiation pressure and assuming a gas of
primordial composition (i.e.\, metal and dust free). We will relax some
of these assumptions in future works. However, the parametric formulas
for the accretion presented in this paper should provide a realistic
description of quasi-spherical accretion onto IMBH for ambient gas
densities $n_{H, \infty} \lesssim 10^5-10^6$~cm$^{-3}$, as radiation
pressure should be minor for these densities. 


We find an oscillatory behavior of the accretion rate that can be
explained by the effect of UV and X-ray photo-heating. The ionizing
photons produced by the black hole near the gravitational radius
increase gas pressure around the black hole. This pressure prevents
the surrounding gas from being accreted. An over-dense shell starts to
form just outside the Str\"{o}mgren sphere. Due to the decreased
accretion rate, the number of emitted ionizing photons decreases and
the density inside the Str\"{o}mgren sphere also decreases with
time. Gas accretion onto the black hole is dominant in decreasing the
density inside the HII region only for ambient gas density $n_{H,
\infty} \gtrsim 10^7$~cm$^{-3}$; for lower values of the ambient gas
density the gas inside the HII region is pushed outward toward the
dense shell by a pressure gradient that develops behind the ionization
front.  Eventually, the pressure gradient inside the Str\"{o}mgren
sphere is not able to support the weight of the over-dense shell that
starts to fall toward the black hole. The accretion rate rapidly
increases and the Str\"{o}mgren sphere starts to expand again.

However, the introduction of a small, non-zero, angular momentum in
the flow could change the time-dependent behavior of accretion and
feedback loop. The inflow rate in the accretion disk that will
necessarily develop, and that is not resolved in our simulations,
is typically much slower than the free-fall rate since the viscous
time scale in units of the free-fall time is $ t_{visc}/t_{ff} \sim
\alpha^{-1} \mathcal{M}^2 $ where $\alpha$ is the dimensionless parameter
for a thin disk \citep*{ShakuraS:1973} and $\mathcal{M}$ is the gas Mach
number. Therefore, angular momentum may produce a long delay between
changes in the accretion rate at the inner boundary of our simulation
and their mirror in terms of output luminosity.  Hence $\alpha \simeq
0.01-0.1$ and $\mathcal{M}$ at the inner boundary of our simulations
is of order of unity, time delays of 10-100 free-fall
times are shorter if the disk is smaller than the inner boundary of the
simulation.  We have started investigating the effect of such a delay
on the periodic oscillations and preliminary
results show that the period of the oscillations can be modified by the
time delay but the oscillatory behavior is still present (at least for
delays of 10-100 free fall times calculated at the inner boundary). 
As long as the time delay is shorter than the oscillation period, that
depends mainly on the gas density, it does not seem to affect the results. 
We are carefully investigating this
in the low and high density regimes where the oscillation pattern and the
periods are different. At low densities a time delay of a few hundred
free-fall times is much smaller compared to the oscillation period,
whereas at the high densities the maximum time delay that we have tested
is comparable to the oscillation period.  We will publish more extensive
results on this effect in our next paper in this series. 

We find that the average accretion rate is sensitive to the
temperature of the ambient medium and to the temperature profile
inside the ionized bubble, and so depends on the gas cooling function
and spectral energy distribution of the radiation. The period of the
accretion bursts is insensitive to the temperature structure of the
HII region, but is proportional to its radius.

Our simulations show that 1D results adequately reproduce 2D results
in which instabilities often develop. We find that the accretion rate is
expressed as $\lambda_{rad} \simeq 3\%~T_{\infty,4}^{2.5} \left(
  T_{in}/4\times 10^4~{\rm K} \right)^{-4}$.  We also derive
$\tau_{cycle}$ as a function of $\eta$, $M_{bh}$, $n_{H,\infty}$ and
$T_{\infty}$. The dependencies of $\langle \lambda_{rad} \rangle$ and
$\tau_{cycle}$ on our free parameters can be explained
analytically. Assuming pressure equilibrium across the Str\"{o}mgren
sphere is a key ingredient to derive the dependence of $\langle
\lambda_{rad} \rangle$ on $T_\infty$, whereas the linear relation
between the average size of the Str\"{o}mgren sphere and
$\tau_{cycle}$ is used to derive the dependence of $\tau_{cycle}$ on
all the parameters we varied.

The qualitative picture of the feedback loop agrees with the
description of X-ray bursters in \cite{CowieOS:78}. After
extrapolating our analytical formulas to black holes of a few solar
masses studied by \cite{CowieOS:78}, we find that the average
accretion rate is in good agreement ($L \sim 2\times
10^{35}$~erg/s). However, the details of the accretion rate as a
function of time, the burst period and peak accretion rates show
qualitative differences. \cite{CowieOS:78} simulations do not show
periodic oscillation while our simulations have well-defined fast rise
and exponential decay of accretion followed by quiescent phases of the
accretion rate. This regular pattern of accretion bursts is possible
only when spherical symmetry is maintained on relatively large scales
during oscillations. An axisymmetric radiation source
\citep{Proga:07,ProgaOK:08,KurosawaPN:09,KurosawaP:09a,KurosawaP:09b}
or inhomogeneous initial condition on scale of the Bondi radius can
break the symmetry.

Our simulations are also in excellent qualitative agreement with
simulations by MCB09 that studied accretion onto $100$~M$_\odot$ black
hole for the case $n_{H,\infty}=10^7$~cm$^{-3}$.  However, we find a
dimensionless accretion rate $\langle \lambda_{rad} \rangle \sim 3\%$
($\langle \lambda_{rad} \rangle \sim 2\%$ including helium
heating/cooling) that is about one order of magnitude larger than in
MCB09.  The cycle period, $\tau_{cycle}$, is in better agreement since
$\tau_{cycle}\propto \langle R_s \rangle \propto
\langle\lambda_{rad}\rangle^{1/3}$. The discrepancy in the mean
accretion is likely produced by the effect radiation pressure on HI,
that becomes important for large $n_{H, \infty}$ and that we have
neglected. In addition, our results indicate that the qualitative
description of the feedback loop starts to change at ambient densities
$>10^6-10^7$~cm$^{-3}$: the oscillation period decreases much more
rapidly with increasing ambient density as gas depletion inside the
ionized bubble becomes dominated by accretion onto the IMBH. The
accretion luminosity during the quiescent phase of the accretion also
increases and the two phases of growth and collapse of the dense shell
become blended into a smoother modulation of the accretion
rate. Hence, further numerical studies are required to characterize
accretion onto IMBH in the high-density regime.

As mentioned above, in this study we have neglected three important
physical processes that may further reduce the accretion rate: 1)
Compton heating, 2) radiation pressure and 3) Lyman-$\alpha$
scattering processes. The importance of these processes is thoroughly
discussed in MBCO09. \citet{Ricotti:08} and MBCO09 find that Compton
heating is not an important feedback mechanism in regulating the
accretion rate onto IMBHs in which the gas density inside the
Str\"{o}mgren sphere is roughly independent of the radius. However,
MBCO09 suggest that both radiation pressure on HI and Lyman-$\alpha$
radiation pressure can contribute to reducing the accretion rate onto
IMBH.
At higher densities accretion becomes Eddington limited
($\langle \lambda_{rad} \rangle \dot{M}_B \lesssim \dot{M}_{Edd}$). 
By assuming $\langle \lambda_{rad} \rangle \sim 1\%~T_{\infty,4}^{2.5}$ which
is suggested by simulations including helium and manipulating Equation
(\ref{eq:edd}) we obtain $M_{BH,2}~T_{\infty,4}~n_{H,5}~\eta_{-1} \gtrsim 40$
as a criteria for Eddington
limited condition. This expression predicts that radiation 
pressure becomes important at gas density 
$n_{H,5} \sim 10-100$ with other parameters fixed to unity. 
It also implies that this critical density depends on the black hole mass, 
gas temperature and radiative efficiency. 
We expect radiation pressure to play a minor role at low densities 
also because the accretion luminosity is negligible during the quiescent phases of accretion (between accretion bursts). 
However, at densities $n_{H, \infty}\ge 10^7$~cm$^{-3}$ the accretion 
rate during the quiescent phases is not negligible, thus radiation pressure
can be important in this regime. The main
effect of Thomson radiation pressure is to prevent the accretion
luminosity to exceed the Eddington limit. The continuum radiation
pressure due to HI ionization can instead be important for
sub-Eddington luminosities, but will be strong only at the location of
the ionization front during the peak of the accretion burst. We will
present more extensive results in the next paper of this series. 

The results of this study provide a first step to estimate the maximum
X-ray luminosity and period of oscillations of an accreting IMBH from
a medium with given physical conditions. Hence, they may be useful for
modeling detection probability of ULX originating from accreting IMBH
in the local universe. From the average growth rate of IMBHs accreting
in this manner it is also possible to estimate the maximum masses of
quasars at a given redshift starting from seed primordial black
holes. One of the main motivations of this study is to derive simple
analytical prescriptions to incorporate growth of seed black holes
from Population~III stars into large scale cosmological
simulation. However, before being able to do use these results in
cosmological simulation we need to understand the effects of relaxing
our assumption of spherical symmetric accretion: we need to simulate
accretion onto moving black holes
\citep{HoyleL:39,Shima:85,Ruffert:94,Ruffert:96} and use more realistic
initial conditions, including gas with non-zero angular momentum or
a multi-phase turbulent ISM \citep{Krumholz:05,Krumholz:06}.

\acknowledgments
The simulations presented in this paper were carried out using high performance 
computing clusters administered by the Center for Theory and Computation of the
Department of Astronomy at the University of Maryland ("yorp"), and the
Office of Information Technology at the University of Maryland ("deepthought").
This research was supported by NASA grants NNX07AH10G and NNX10AH10G. The
authors thank the anonymous referee for constructive comments and
feedback.

\bibliographystyle{apj}
\bibliography{pr10_revIII}




\appendix

\section{BASIC TESTS OF THE CODE}
\begin{figure*}[thb]
\epsscale{1.0} \plottwo{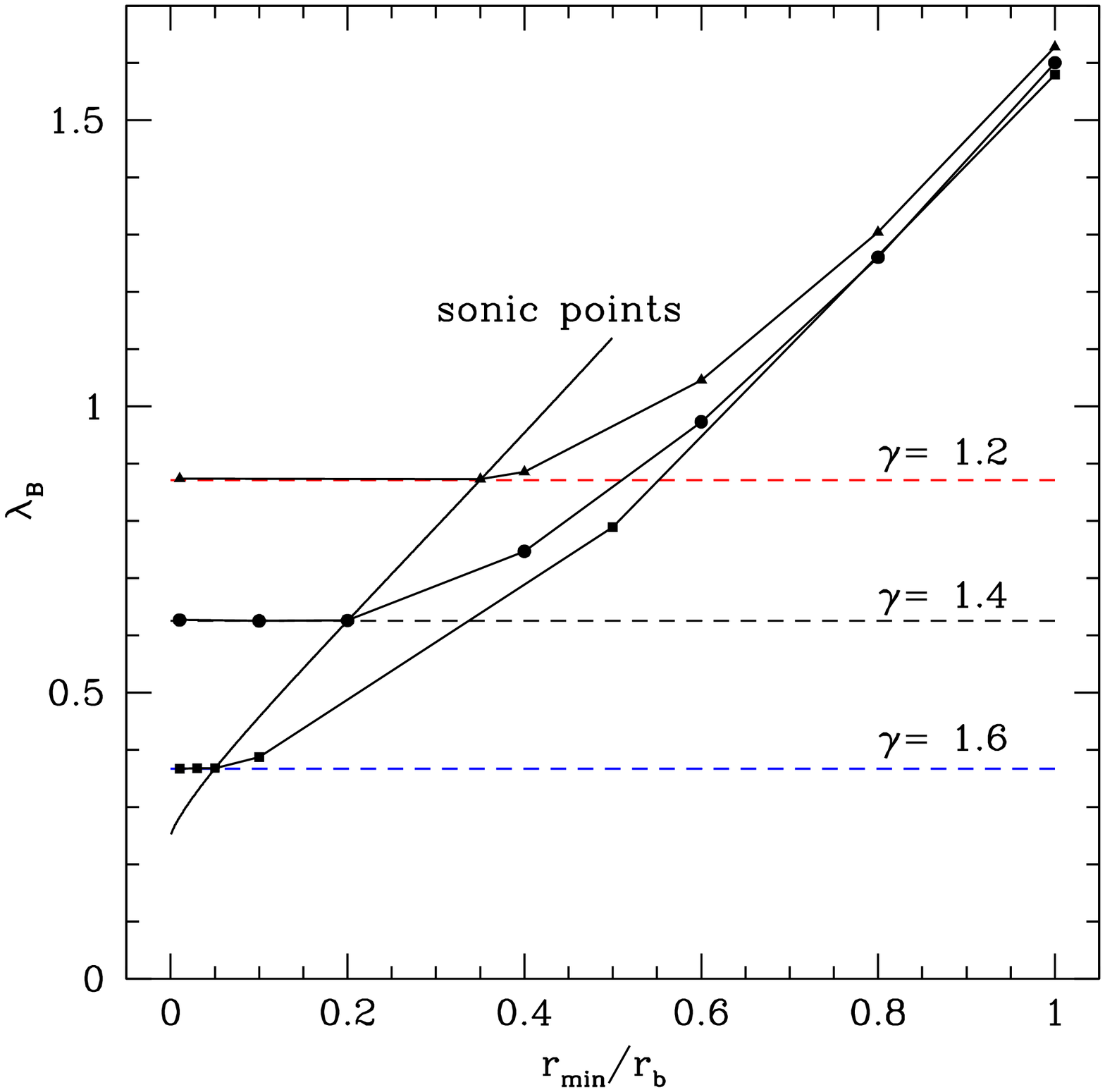}{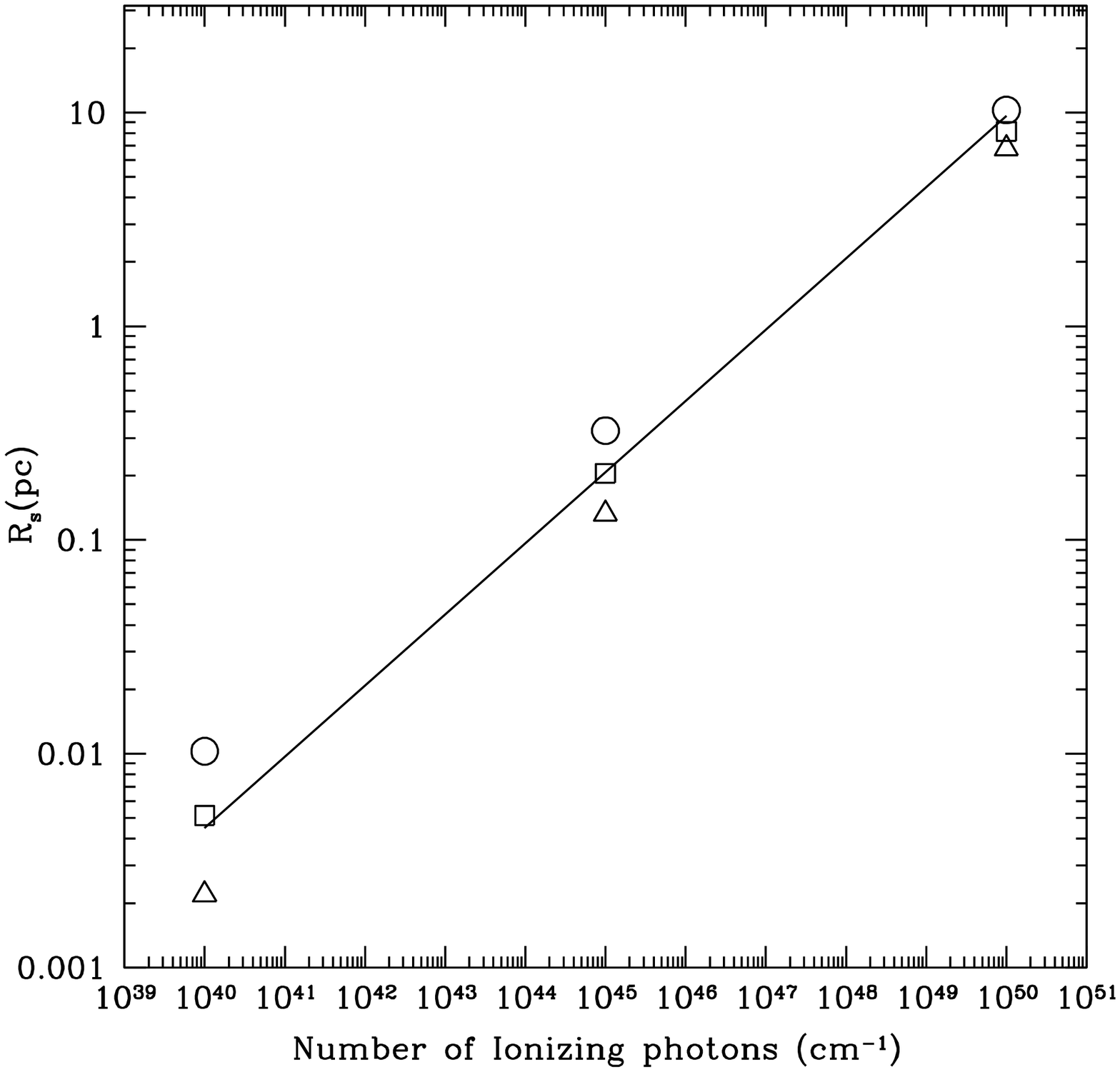}
\caption{ {\it Left} : Simulated Bondi accretion rate($\lambda_{B}$)
  as a function of minimum radius with given adiabatic index $\gamma$
  without radiative feedback. Dashed lines are analytically estimated
  values for each $\gamma =$1.2, 1.4 and 1.6. In order not to
  overestimate accretion rate sonic point should be resolved where the
  velocity of the inflowing gas becomes supersonic.{\it Right} : Test
  of Str\"{o}mgren radius with given number of ionizing photons. Solid
  line is the prediction for the given number of ionizing photons from
  $10^{40}$ to $10^{50}~$s$^{-1}$. Triangle symbols represent location
  where ionization fraction of $n_{\rm H}$ ($x_{\rm HI}$) is
  0.50. Squares are for $x_{\rm HI} = 0.90$ and circles are for
  $x_{\rm HI}=0.99$.}
\label{acc_gamm}
\end{figure*}

We test the Bondi accretion formula using ZEUS-MP for the adiabatic
indexes $\gamma =1.2, 1.4$ and $1.6$. For a given equation of state,
the sonic point where the gas inflow becomes supersonic must be
resolved not to overestimate the accretion rate $\lambda_{B}$. The
left panel of Figure \ref{acc_gamm} shows the steady accretion rate as
a function of the radius at the inner boundary normalized by the Bondi
radius. Different lines show results for $\gamma=$1.2, 1.4 and 1.6.

We also test whether our radiative transfer module produces radii of
the Str\"{o}mgren spheres in agreement with the analytical prediction:
$(4\pi/3)R_s^3 n_e n_H \alpha_{rec} = N_{ion}$, where $R_s$ is the
Str\"{o}mgren radius and $N_{ion}$ is the number of ionizing photons
emitted per unit time. The right panel of Figure \ref{acc_gamm} shows
the test of the 1D radiative transfer module without
hydrodynamics. Different symbols indicates the radii for the different
ionization fractions: $x_e =$ 0.99 (circle), 0.90 (square), 0.50
(triangle).

\section{RADIATIVE TRANSFER MODULE AND TIME STEPPING}
Our hydrodynamic calculation is performed using ZEUS-MP, returning the
density and gas energy at each time step to the radiative transfer
module. The operator-splitting method is applied to mediate between
hydrodynamics and radiative transfer with a photon-conserving method
\citep{Whalen:06}. For each line of sight radiative transfer
equations are solved in the following order:

\begin{enumerate}
\setlength{\itemsep}{1pt}
\setlength{\parskip}{0pt}
\setlength{\parsep}{0pt}

\item At the inner boundary, the average inflow mass flux $\dot{M}$ is
  calculated.
\item The mass flux is converted into accretion luminosity $L$, and
  thus into the number of ionizing photons for a given radiative
  efficiency $\eta$.
\item The photon spectrum is determined using a power law spectral
  energy distribution with the spectral index $\alpha$. We use up to
  $300$ logarithmically spaced frequency bins for photons between
  $10$~eV up to $100$~keV.
\item The ordinary differential equation for time-dependent radiative
  transfer cooling/heating and chemistry of the gas are solved using a
  Runge-Kutta or Semi-Implicit solver for each line of sight with a
  maximum of 10\% error. Photo-heating, cooling for a given cooling
  function and Compton cooling are calculated.
\item The energy density and the abundances of neutral and ionized
  hydrogen are updated.
\end{enumerate}
Parallelization is easily implemented in polar angle direction because
radiative transfer calculations along each ray are independent of one
another.

\section{RESOLUTION STUDIES}
\begin{figure}[thb]
\epsscale{0.5} \plotone{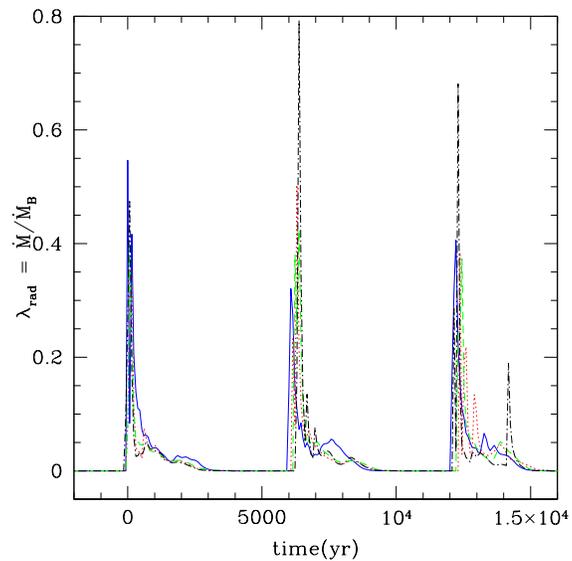}
\caption{Comparisons between simulations of $\eta = 0.1$,
  $M_{bh}=100$~M$_{\sun}$, $n_{H,\infty}=10^{5}$~cm$^{-3}$ and
  $T_{\infty}= 10^4$~K with various resolution. {\it Solid} : 384 grid
  run. {\it Dotted} : 512 grid run. {\it Long dashed} 768 grid
  run. {\it Short dashed} : 512 grid with Courant number of 0.05.}
\label{strom}
\end{figure}

We perform a resolution study to confirm that the number of grid zones
does not affect the results. Number of zones from 384 to 768 are
tested and they all show the similar outputs in terms of accretion
rate at peaks, average accretion rate, decaying shape and the period
between peaks . Figure \ref{strom} shows that the details of the
accretion rate history from simulations are not identical but the
physical quantities which we are interested in (average accretion
rate, peak accretion rate and period of the bursts) do not show
significant deviation from each other. In general, a Courant number of
0.5 is used for most simulations, but we try a Courant number which is
one order of magnitude smaller to investigate how the results are
affected by reducing the hydro-time step by an order of magnitude. The
chemical/cooling time steps are calculated independently by the
radiation transfer module. 

\end{document}